# Unified description of the proton, alpha, cluster decays and spontaneously fissions half- life

**Strachimir Cht. Mavrodiev[1], M.A. Deliyergiyev[2]**

[1] Institute for Nuclear Researches and Nuclear Energy, BAS, Sofia, Bulgaria
[2] Department of high energy Nuclear Physics, Institute of Modern Physics, CAS, Lanzhou, China

**Abstract**

Some time ago there was demonstrated the possibility of classical (without Gamow tunneling) universal description of radioactive nuclei decay. Such possibility is bases on the classical interpretation of Bohmian Psi – field reality in Bohmian- Chetaev mechanics and the hypothesis for the presence of dissipative forces, generated from the Gryzinsky translational precession of the charged particles spin, in Langevin- Kramers diffusion mechanism.

In this paper is present an unified model of proton, alpha decay, cluster radioactivity and spontaneous fission half- time as explicit function which depends on the total decay energy $Q_t$ and kinetic $E_k$ energy, the number of protons $Z_{cl}$ and neutrons $N_{cl}$ of daughter product, the number of protons Z and neutrons of mother nuclei and from a set of **a = ($a_i$, i=1,…,n)** unknown digital parameters. The Half- lifes of the 573 nuclei taken from NuDat database together with the recent experimental data from Oganessian provide a basis for discovering the explicit form of the Kramers's solution of Langevin type equation in a framework of inverse problems with the help of the Alexandrov's dynamic auto-regularization method (FORTRAN program REGN-Dubna). The procedure LCH in program REGN permitted to reduce the number of unknown digital parameters from 137 to 79. The model describes 424 decays quantities with deviation of order one in year's power scale.

1. ## Introduction

The beginning of the twentieth century brought surprising non-classical phenomena. Max Planck's explanation of the black body radiation [1], the work of Albert Einstein on the photoelectric effect [2], the Niels Bohr's model of the electron orbits around the nuclei [3], the existence of protons and neutrons in the atomic nuclei [4,5], established what is now known as a quantum theory.

### 1.1. Bohmian mechanics.

The ignored by the scientific community till present Bohmian quantum mechanics was first proposed by Louis de Broglie [6] and rediscovered by David Bohm [7] many years later.

In 1926 Schrodinger published his equation for the wave function (field) *ψ(r, t)* [8]. In 1932 von Neumann put quantum theory on rigorous mathematical basis [9]. The main result was a state that quantum-mechanical probabilities cannot be understanding in terms of any conceivable distribution over hidden parameters.

In 1935 Einstein, Podolsky and Rosen [10], based on the hypothesis for the absence of action at a distance, argued that the quantum theory is either nonlocal or incomplete.

In 1952 David Bohm [7] demonstrated that von Neumann theorem [8] has limited validity.

In 1964, inspired by the EPR paper [10] and Bohm's works on nonlocal hidden variables [7], Bell elaborated a theorem establishing clear mathematical inequalities, now known as Bell inequalities, for experimental results that would be fulfilled by local theories but would be violated by nonlocal ones [11]. In 1987 Bell explained that the orthodox (Copenhagen) interpretation of the Quantum mechanics is not adequate to the processes in the Nature, nevertheless that the calculations describe the experimental data [12].

Contemporary clear and full presentation of Bohmian quantum mechanics, including its chemical many particles applications, which can be seen in papers [14, 15]

**1.2. The Chetaev's stable motion**

In 30-th of 20-th century Nikolai Gurevich Chetaev [16], used the Lyapunov theorem for arbitrary small perturbation forces, which can do the motion unstable, formulated his famous theorem for "stable trajectories in dynamics". The reason for such stability Chetaev explained with existence of small dissipative forces with full dissipation, which always exist in the nature. Analyzed a holonomic mechanical systems, Chetaev demonstrated that its solutions give us a picture of quantum phenomena, because of analogy with Schrodinger type equation.

The origin of dissipative forces for the stable movement (orbits) of the electrons in the atoms can be the precession of the proton and electron spin. Such statement was proposed in by Michal Gryzinsky [18], who demonstrated the description of Hydrogen atom Ballmer's series in a frame work of classical Kepler problem with phenomenological vector potential. The source of the Gryzinsky potential is the Coulomb interaction and the oscillating electromagnetic field of photon or electron caused by translational precession of the spin

**2. The Langevin- Kramers description of the nucleus decays and fissions**

In papers [19] the Chetaev theorem on stable trajectories in dynamics was generalized to the case when the Hamiltonian of a system is explicitly time- dependent. In a case of particle with mass $m$ in the field of conservative forces presented by $U$, which depends on the time the result was that the Chetaev's motion stability condition has the form of Schrodinger equation

$$i\hbar\frac{\partial \Psi}{\partial t} = -\frac{\hbar^2}{2m}\Delta\Psi + U\Psi \qquad (1)$$

The substitution $\Psi = Ae^{\frac{i}{\hbar}S}$, where S is the classical action, in eq. (1) born an equivalent 2-system of equations, known as Bohm- Madelung system of equations [14], [20]:

$$\frac{\partial A}{\partial t} = -\frac{1}{2m}(A\,\Delta S + 2\,\nabla A.\nabla S).$$

(2)

$$\frac{\partial S}{\partial t} = -[\frac{(\nabla S)^2}{2m} + U - \frac{\hbar^2}{2m}\frac{\Delta A}{A}].$$

It is important to note that the last term in eqs. (2)

$$Q = -\frac{\hbar^2}{2m}\frac{\Delta A}{A}$$

is the quantum potential of Bohm $\Psi - field$.

We have, after the substitution $P(q,t) = \Psi\Psi^* = A(q,t)^2$ in eqs. 2 the forms

$$\frac{\partial P}{\partial t} = -\frac{1}{m}\nabla(P.\nabla S) \qquad (3)$$

and

$$\frac{\partial S}{\partial t} + \frac{(\nabla S)^2}{2m} + U - \frac{\hbar^2}{4m}[\frac{\Delta P}{P} - \frac{1}{2}\frac{(\nabla P)^2}{P^2}] = 0. \qquad (4)$$

The solution of eq. (3) is the probability density $P(q,t)$ to find the particle from eq. (1) in a certain point in space- time. According eq. (4) the vector variable $\mathbf{v} = \frac{\nabla S}{m}$ has sense of velocity.

In papers [21] for describing the half- life data of the alpha, cluster decays and spontaneous fissions was use the Kramers diffusion mechanism over potential barrier [22]. The theoretical argument was that the probability density $P(q,t)$ moves according to the laws of classical mechanics with a classical velocity $\mathbf{v} = \frac{\nabla S}{m}$ [19].

The explicit form of Half-life formulae was derived in the frame work of Langevin theory of Brownian motion [23], Kramers diffusion mechanism over potential barrier [22] and Fermi- gas model for connection between thermodynamic temperature and internal excitation energy of many particles system [24]:

$$\lg T_{1/2} = -\lg\frac{w_{Kramers}}{2\pi} + \lg(\exp)\left(\frac{A}{8\mu}\right)^{1/2}\frac{V_{Coul}-E_k}{\sqrt{E_k}}, \qquad (5)$$

where the charge Coulomb potential is

$$V_{Coul} = \frac{(Z-Z_{cl})}{R_{Coul}}Z_{cl}$$

and

$$R_{Coul} = R_{A-A_{cl},Z-Z_{cl}} + R_{A_{cl},Z_{cl}} + R_{NuclF},,$$

*A* and *Z* are the mass number and charge of parent nucleus, $Z - Z_{Cl}$ is the charge of the daughter nucleus and $R_{Coul}$ [fm] is the minimal Coulomb radius and the variable $E_k$ **[MeV]** is kinetic energy of the cluster.

It is important to note that from numerical point of view (the number of freedom in sense of Weierstrass approximation theorem) the half-times according "Kramers's over potential barrier picture" and "Gamow's tunneling quantum mechanism [25]", are equivalent to the Geiger-Nuttall formulae [26].

The aim of this paper is to apply the results of paper [21] for description the half- life data for of proton (22), alpha (497), cluster (28) decays and spontaneous fissions (26) from NUDAT data- base [27] and paper [28], total 573 data.

In Application A is presented the FORTRAN code of function *Ht1/2(Z, N, $Z_{cl}$ , $N_{cl}$ , $E_k$ , $Q_t$, a)*. In application B is presented the result table for described and experimental half-times.

### 3. The formulation of inverse problem

If we rewrite the formulae (5) in form

$$T^{Th}_{1/2}(\mathbf{Z}, \mathbf{N}, \mathbf{a}) = 10^{(\mathcal{W}(\mathbf{Z},\mathbf{N},\mathbf{a}) + \mathcal{U}(\mathbf{Z},\mathbf{N},\mathbf{a}))} \quad , \tag{6}$$

where $\mathcal{W}(\mathbf{Z}, \mathbf{N}, \mathbf{a})$ and $\mathcal{U}(\mathbf{Z}, \mathbf{N}, \mathbf{a})$ are unknowns functions of variables *Z, N, $Z_{cl}$, $N_{cl}$,* kinetic energy $E_k$, total energy $Q_t$, isotopic spin characteristics of parent nuclei and set **a** = *{$a_i$}* = **a** $_i$, *i=1,…,n*) unknown digital parameters, the solution of over determined nonlinear system of **M** algebraic equations for **n** real unknowns parameters:

$$T^{Expt}_{1/2}(Z_j, N_j) = T^{Th}_{1/2}(Z_j, N_j, ((a_i), i = 1, ... n)) , j=1,…,M , \tag{7}$$

where *M≥ n,* where ***Expt*** and ***Th*** means experiment and model correspondingly**.**

**3.1. Solution of the overdetermined system of equations and discovering the explicit form of functions**

For solution of ill-posed problem [29] (7) we use the Alexandrov dynamic autoregularization method (FORTRAN code REGN-Dubna [30- 32]. The use of procedure LCH in REGN permits to discover the explicit form of unknown function because one can chose uniquely the better from two functions which use gives the same hi- squared $\chi^2$ [33].

### 4. Unified description of the proton, alpha, cluster decays and spontaneously fissions half time

In NuDat database [27] the 519 half-time data for proton, alpha, cluster decays and spontaneous fissions are published. Also 54 data for the alpha decay are published in [28]. So, in our database we have 573 half- time data.

**4.1. The choose of the variables**

For solving such type of inverse problems it is very convenient the chosen variables to be in the interval +/-1 as well as the variables to be linearly independent.

By using simple nonlinear substitutions from 8 variables A = Z + N, Z, N, N - Z, $A_{cl}$ = $Z_{cl}$ + $N_{cl}$, $Z_{cl}$, $N_{cl}$, $N_{cl}$ - $Z_{cl}$ we create the variables:

$$v_1 = \frac{Z}{A},\ v_2 = \frac{N}{A},\ v_3 = \frac{N-Z}{A},\ v_4 = \frac{Z}{A_{cl}},\ v_5 = \frac{N_{cl}}{A_{cl}},\ v_6 = \frac{N_{cl}-Z_{cl}}{A_{cl}},\ v_7 = \frac{E_k}{Q_t},\ v_8 = \frac{Z_{cl}(Z-Z_{cl})}{Z\,Z_{cl}}.$$

The isotopic spin dependence is included in the analysis with using the variables $v_9 = v_{10} = v_{11} = 0$, if A, Z, N are even and if A, Z, N are odd $v_9 = v_{10} = v_{11} = 1$.

### 4.2. Explicit form of the half- time as function of Z, N and parameters a

The explicit from of the function $\mathcal{W}(Z, N, a)$ is:

$$\mathcal{W}(Z, N, a) = -e^{a_{n-2}} R(Z, N, a) + \mathrm{MagNum}_c(Z, N, a).$$

$$\mathrm{MagNum}_c(Z, N, a) = A_Z(Z, N, a) \frac{\exp(\frac{(Z-Z_{MN})^2}{w_Z^2})}{((Z-Z_{MN})^2 + w_Z^2)} + A_N(Z, N, a) \frac{\exp(\frac{(N-N_{MN})^2}{w_N^2})}{((N-N_{MN})^2 + w_N^2)},$$

$$A_Z(Z, N, a) = e^{\left(a_{Np2+4} + \sum_{i=1}^{Np1}(a_{i+9.Np1} v_i + a_{i+10.Np1} v_i^2 + a_{i+11.Np1} v_i^3)\right)},$$

$$A_N(Z, N, a) = e^{\left(a_{Np2+5} + \sum_{i=1}^{Np1}(a_{i+12.Np1} v_i + a_{i+13.Np1} v_i^2 + a_{i+14.Np1} v_i^3)\right)},$$

$Z_{MN}, N_{MN}$ are the nearest to Z and N magic numbers and $w_Z, w_N$ are equal to the half of difference between magic numbers in which interval belong Z and N correspondingly.

For the radius [21] $R(Z, N, a)$ with influence of isotopic spin correction we have:

$$R(Z, N, a) = \left(Be(Z, N, a)\left((A - A_{cl})^{\frac{1}{3}} + A_{cl}^{\frac{1}{3}}\right) - 1\right) Ce(Z, N, a),$$

$$Be(Z, N, a) = e^{\left(a_{Np2+2} + \sum_{i=1}^{Np1}(a_{i+3.Np1} v_i + a_{i+4.Np1} v_i^2 + a_{i+5.Np1} v_i^3) + B_c(Z,N,a)\right)},$$

$$B_c(Z, N, a) = \sum_{i=1}^{3} a_{Ns+3+i}\, v_{8+i}$$

$$Ce(Z, N, a) = e^{\left(a_{Np2+3} + \sum_{i=1}^{Np1}(a_{i+6.Np1} v_i a_{i+7.Np1} v_i^2 + a_{i+8.Np1} v_i^3) + C_c(Z,N,a)\right)},$$

$$C_c(Z, N, a) = \sum_{i=1}^{3} a_{Ns+6+i}\, v_{8+i}$$

For the term $\mathcal{U}(Z, N, a)$ in (6) we have [21]:

$$\mathcal{U}(Z, N, a) = \sqrt{\frac{A}{Mu(Z,N,a)}} \; \frac{\frac{Z_{cl}(Z-Z_{cl})}{R\,(Z,N,a)} - E_k}{\sqrt{E_k}},$$

$$Mu(Z, N, a) = a_{n-1}\left(1 + e^{-\left(a_{Np2+1}+sMu(Z,N,a)\right)^2}\right).$$

$$sMu(Z, N, a) = \sum_{i=1}^{Np1}(a_i v_i + a_{i+Np1}v_i^2 + a_{i+2.Np1}v_i^3) + Mu_c(Z, N, a),$$

$$Mu_c(Z, N, a) = \sum_{i=1}^{3} a_{Ns+6+i}\, v_{8+i}.$$

The integers in the above formulae have a values *Np1 = 8, Nc = 3, Np2 = 15 Np1 = 120, Ns = Np2 + 5 = 125, n = Np2 + 5 + 3 Nc +3 = 137.*

As one can see from Appendix A, where the FORTRAN code of Half-time function is presented the initial number of unknowns parameters *n=137* is reduced to *79*.

### 4.3. The Figures of R (Z, N, a), $\text{MagNum}_c(Z, N, a)$ and $Mu(Z, N, a)$ like functions of $E_k$, Z, N

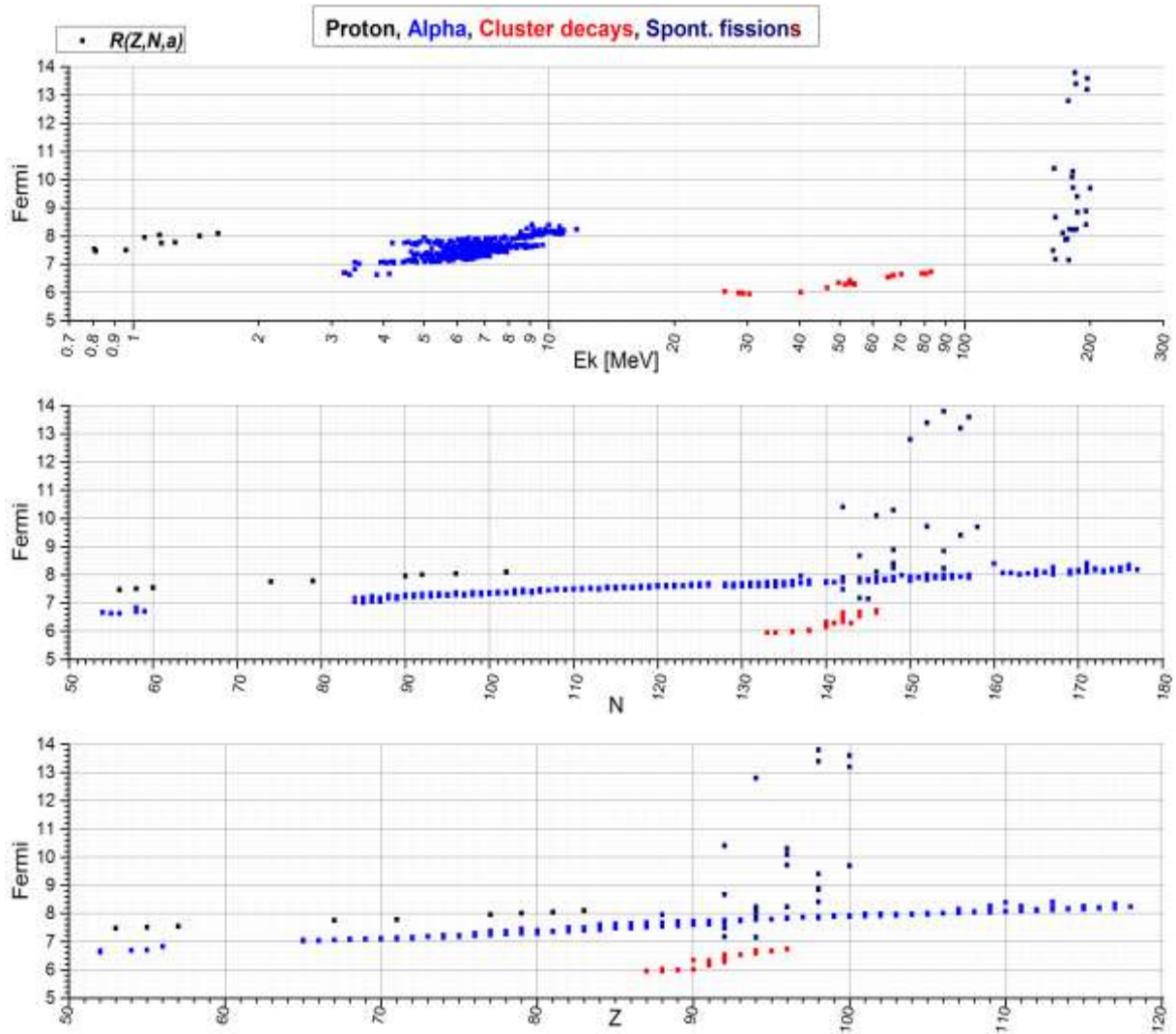

**Fig.1**. $R\ (Z, N, a)$ as function of $E_k$, Z, N

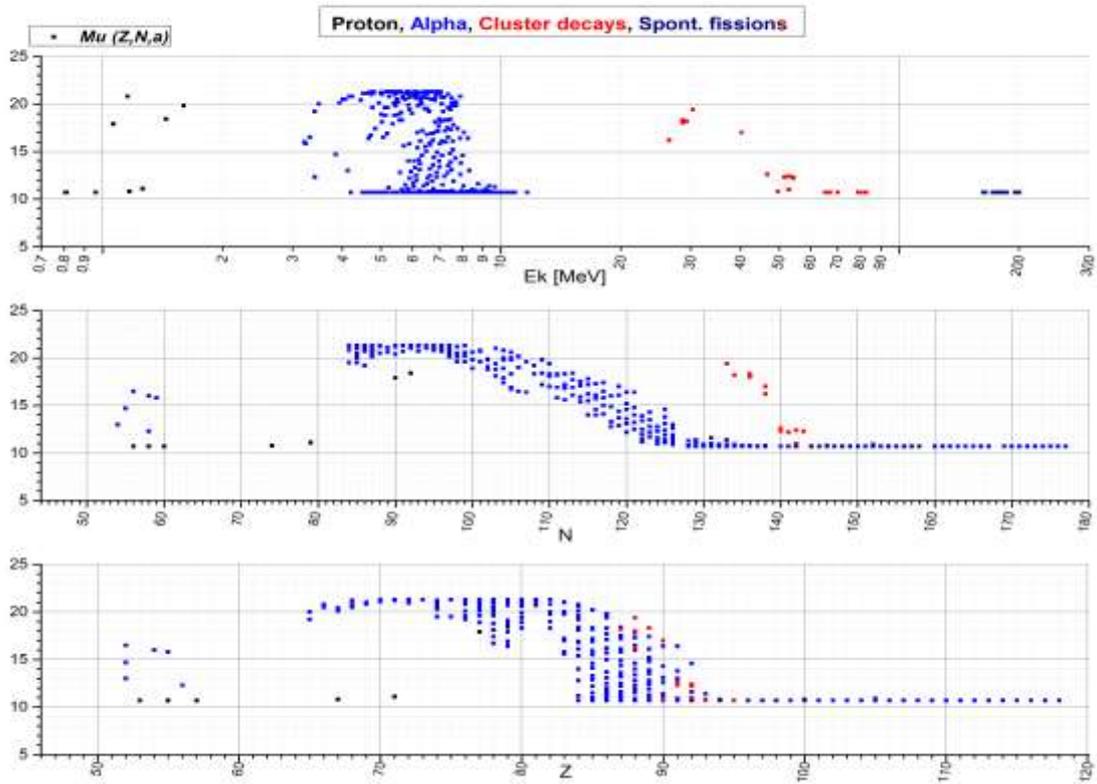

**Fig.2**. $Mu(Z, N, a)$ as function of $E_k$, Z, N

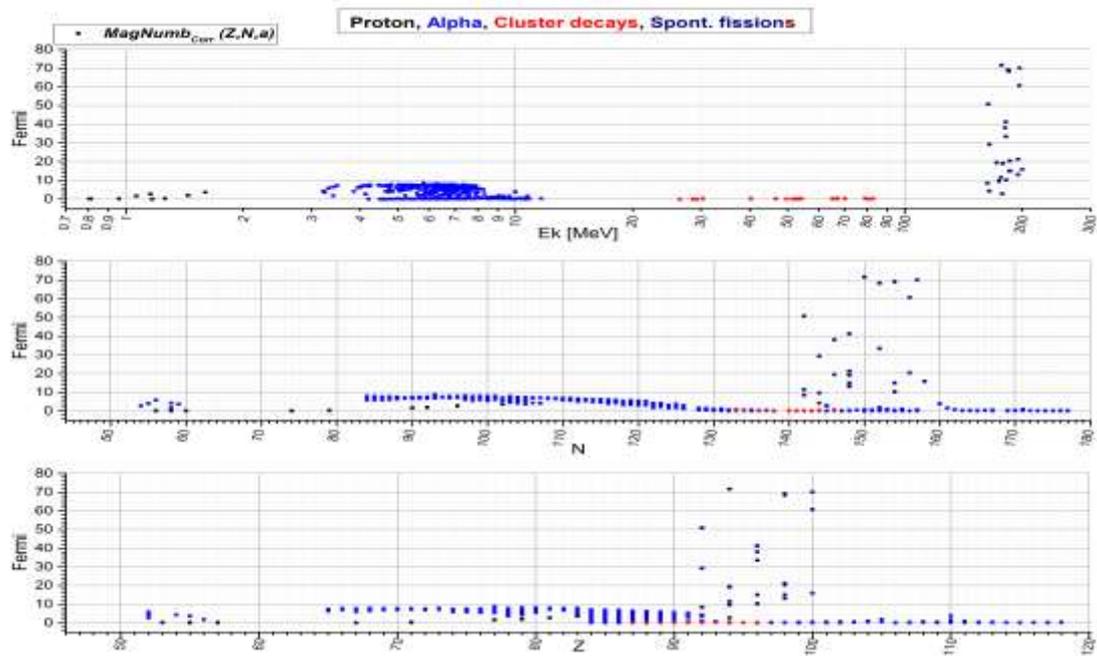

**Fig.3.** $MagNumb_{Corr}(Z, N, a)$ as function of $E_k$, Z, N

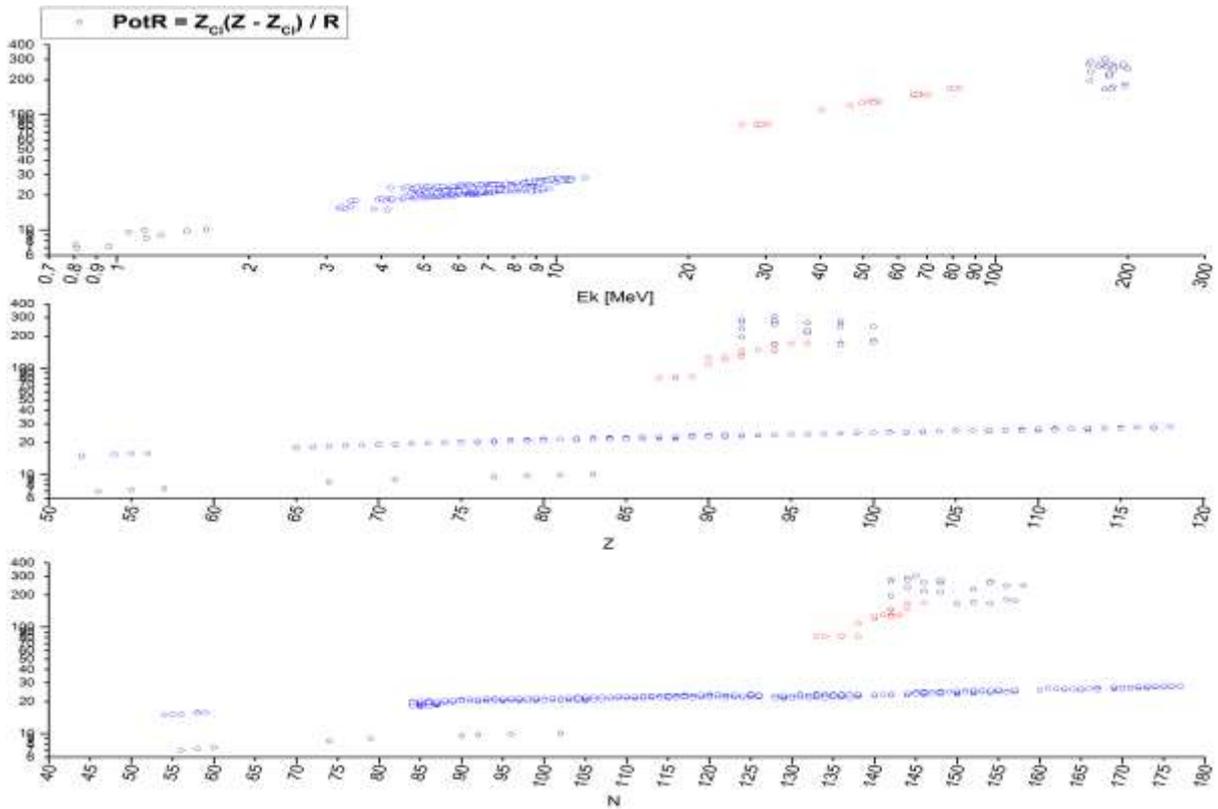

**Fig.4.** *The dependences of* $\frac{Z_{cl}(Z-Z_{cl})}{R\,(Z,N,a)}$ *on variables* $E_k$, Z, N.

## 5. Results

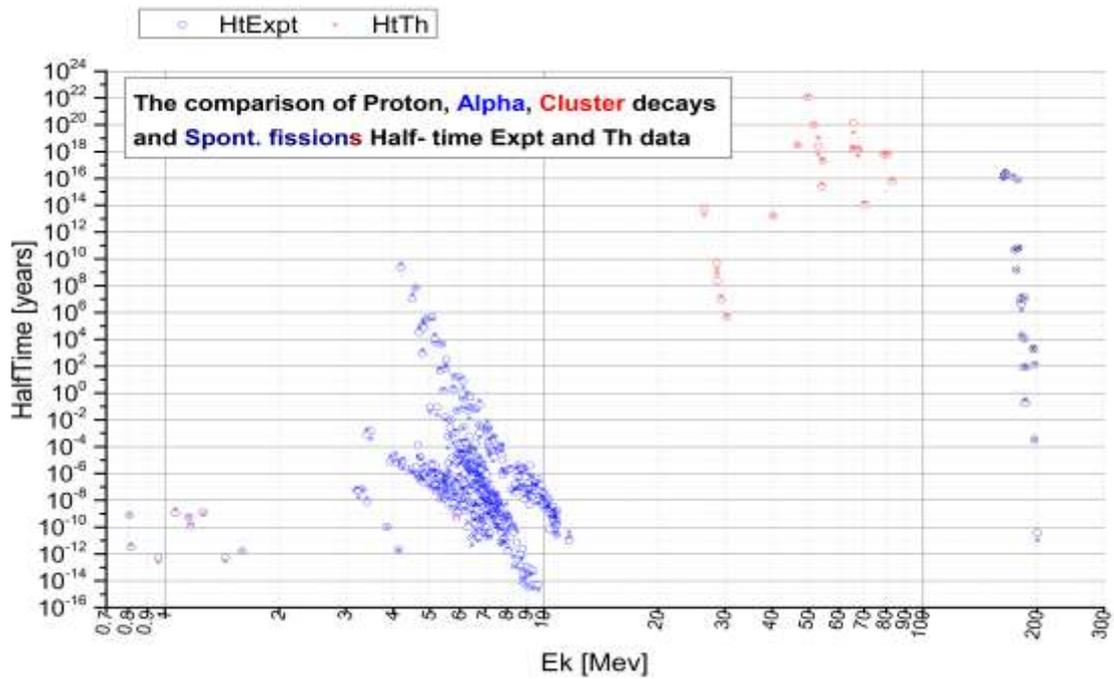

**Fig.4. The comparison of experimental and model data.**

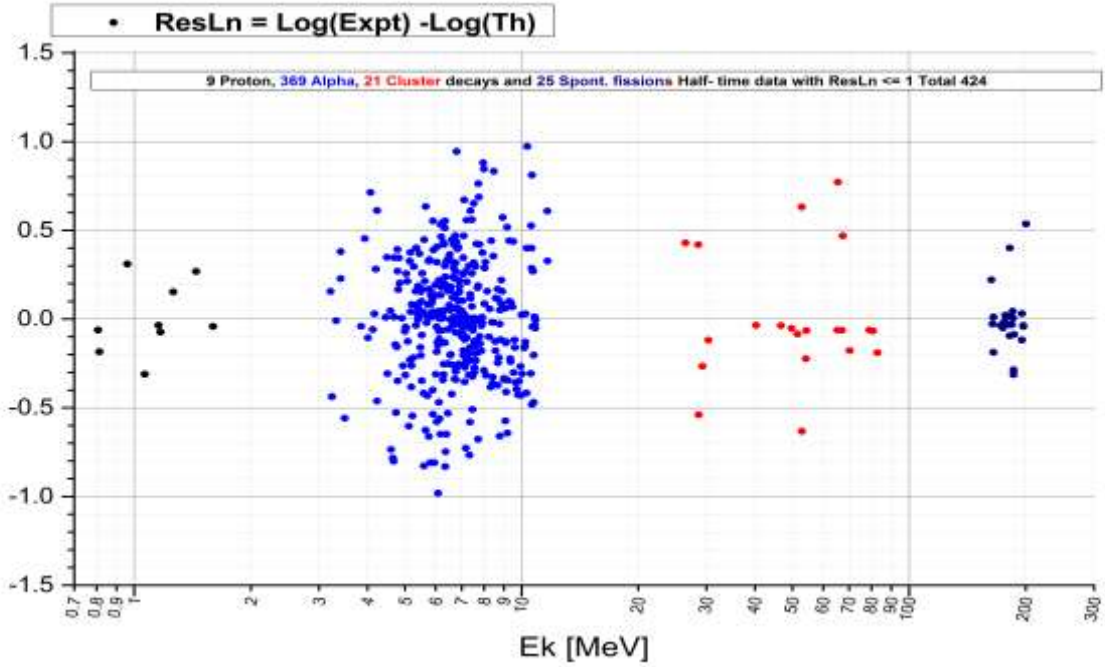

**Fig.5.** The difference between described and experimental values for 10-logaritm of half-lifes with respect to the kinetic energy of the corresponding cluster.

The estimations of description accuracy $\chi^2 = 346$, calculated as follow

$$\chi^2 = \sum_{k=1}^{424} \left( \frac{\text{Expt}(Z_k, N_k) - \text{Th}(Z_k, N_k, a)}{\sigma(Z_k, N_k)} \right)^2,$$

where

$$\sigma(Z_k, N_k) = \sigma_{\text{Stat}}(Z_k, N_k) + \text{Expt}(Z_k, N_k)$$

and $\chi_n = \sqrt{\frac{\chi^2}{346-79}} = \mathbf{1.02}$.

### Conclusion

The presented model of proton, alpha, cluster decays and spontaneous fissions half-time describes with accuracy lest than of one order in year's power scale 9 proton, 368 alpha, 21 cluster decays and 25 spontaneous fissions, total 424 data from 573 initial, as explicit function of the total decay energy $Q_t$ and kinetic $E_k$ energy, the number of protons $Z_{cl}$ and neutrons $N_{cl}$ of daughter product, the number of protons Z and neutrons of mother nuclei and from a set of $a = (a_i, i=1,\ldots,79)$ digital parameters: $Ht1/2(Z, N, Z_{cl}, N_{cl}, E_k, Q_t, a)$.

The result of this paper can be applied in the theoretical researching of stability islands problem (see for example the paper [28]) because the kinetic and total energy of decay $E_k$, $Q_t$ can be calculated using the phenomenological model for nuclei masses [35, 36].

The not accuracy describing if 149 Half- life data from [27,28] data base is, probably, connected with the fact that in our model was not use the information about the eccentricity of the nuclei. In the next paper this problem will be analyzed.

ACKNOWLEDGMENTS: Authors are thankful to Svetla Drenska for many constructive discussions.

## Application 1: The FORTRAN code of Function Ht1/2(Z, N, $Z_c$, $N_c$, $E_k$, $Q_t$, a).

```fortran
      Function HalfTime(Zm,aNm,Zc,aNc,Ek,Qt)
      IMPLICIT DOUBLE PRECISION(A-H,O-Z)
      COMMON/HELP/pi2,Pot,PotR,Te1,Te2,R,Amm,Pow,CorMagicNumbers
      COMMON/nHelp/Np1,Np2,Ns
      DIMENSION A(137), v(11)
A(  1)=  0.00000000000000E+00; A(  2)=  0.00000000000000E+00; A(  3)=  0.00000000000000E+00; A(  4)=  0.00000000000000E+00; A(  5)=  0.36246930734672E+03;
A(  6)=  0.00000000000000E+00; A(  7)=  0.00000000000000E+00; A(  8)=  0.00000000000000E+00; A(  9)= -0.82229629128498E+03; A( 10)= -0.68358201755439E+03;
A( 11)=  0.00000000000000E+00; A( 12)= -0.51964754659017E+03; A( 13)=  0.00000000000000E+00; A( 14)= -0.15460688332192E+03; A( 15)= -0.75416130195326E+02;
A( 16)=  0.14279850650704E+03; A( 17)=  0.00000000000000E+00; A( 18)=  0.00000000000000E+00; A( 19)=  0.11189158516248E+04; A( 20)=  0.00000000000000E+00;
A( 21)= -0.11279073388773E+04; A( 22)= -0.15827657811068E+03; A( 23)=  0.52710307684978E+02; A( 24)=  0.00000000000000E+00; A( 25)=  0.16517683145626E+04;
A( 26)=  0.61977940341912E+03; A( 27)=  0.18858770186569E+03; A( 28)=  0.62698336427981E+03; A( 29)= -0.89384889492166E+01; A( 30)= -0.32245094200614E+01;
A( 31)= -0.16132375325047E+03; A( 32)=  0.31980040070703E+03; A( 33)=  0.69909739831250E+03; A( 34)=  0.85479622859136E+03; A( 35)= -0.55513343492858E+03;
A( 36)=  0.00000000000000E+00; A( 37)= -0.37202319960121E+02; A( 38)=  0.10886592910845E+02; A( 39)=  0.16774174041863E+03; A( 40)=  0.30771482006480E+03;
A( 41)=  0.00000000000000E+00; A( 42)=  0.63227003705520E+03; A( 43)= -0.22534809990904E+03; A( 44)= -0.10381339705593E+03; A( 45)=  0.68211220789148E+02;
A( 46)=  0.74711807307366E+01; A( 47)= -0.57943644153748E+02; A( 48)= -0.10671162839518E+03; A( 49)= -0.24378946942204E+02; A( 50)=  0.00000000000000E+00;
A( 51)=  0.00000000000000E+00; A( 52)=  0.00000000000000E+00; A( 53)=  0.00000000000000E+00; A( 54)=  0.00000000000000E+00; A( 55)=  0.39059920246535E+03;
A( 56)=  0.00000000000000E+00; A( 57)=  0.00000000000000E+00; A( 58)=  0.00000000000000E+00; A( 59)= -0.69423540433091E+02; A( 60)=  0.00000000000000E+00;
A( 61)=  0.10460175139847E+03; A( 62)= -0.66244313006074E+01; A( 63)= -0.41524698628180E+03; A( 64)=  0.58765097365109E+02; A( 65)=  0.00000000000000E+00;
A( 66)=  0.00000000000000E+00; A( 67)=  0.14225405217928E+03; A( 68)=  0.32441703506852E+02; A( 69)= -0.10606140561073E+03; A( 70)=  0.00000000000000E+00;
A( 71)=  0.14675513953838E+03; A( 72)= -0.38585249821046E+02; A( 73)=  0.00000000000000E+00; A( 74)=  0.00000000000000E+00; A( 75)=  0.00000000000000E+00;
A( 76)=  0.00000000000000E+00; A( 77)=  0.76486560088496E+02; A( 78)=  0.00000000000000E+00; A( 79)=  0.00000000000000E+00; A( 80)=  0.00000000000000E+00;
A( 81)=  0.00000000000000E+00; A( 82)=  0.00000000000000E+00; A( 83)=  0.00000000000000E+00; A( 84)= -0.95789785486082E+02; A( 85)= -0.10196340396227E+03;
A( 86)=  0.00000000000000E+00; A( 87)= -0.96173584118981E+02; A( 88)= -0.22073967198270E+03; A( 89)= -0.13155541290215E+03; A( 90)=  0.00000000000000E+00;
A( 91)= -0.96216186528572E+03; A( 92)=  0.00000000000000E+00; A( 93)=  0.00000000000000E+00; A( 94)=  0.00000000000000E+00; A( 95)=  0.65319994758523E+02;
A( 96)=  0.17295387415190E+03; A( 97)=  0.00000000000000E+00; A( 98)=  0.31650140758111E+02; A( 99)=  0.00000000000000E+00; A(100)= -0.11753494293669E+04;
A(101)= -0.84388733012853E+03; A(102)= -0.12429823734332E+03; A(103)= -0.11153534945129E+03; A(104)=  0.00000000000000E+00; A(105)=  0.00000000000000E+00;
A(106)=  0.00000000000000E+00; A(107)= -0.32970243220728E+03; A(108)=  0.00000000000000E+00; A(109)=  0.16867226743925E+04; A(110)=  0.30832142793959E+03;
A(111)=  0.00000000000000E+00; A(112)= -0.98511044293964E+03; A(113)=  0.00000000000000E+00; A(114)=  0.00000000000000E+00; A(115)=  0.41772208091016E+03;
A(116)=  0.00000000000000E+00; A(117)= -0.11694008236937E+04; A(118)=  0.45264988650012E+03; A(119)=  0.39219536033878E+02; A(120)=  0.00000000000000E+00;
A(121)=  0.25229445073453E+03; A(122)= -0.20715089473306E+04; A(123)= -0.13105770018320E+03; A(124)=  0.98933652154999E+02; A(125)=  0.10478516727008E+04;
A(126)=  0.13453892736635E-01; A(127)=  0.21222651017761E-01; A(128)= -0.16218867046091E-01; A(129)= -0.14218009919827E-01; A(130)=  0.00000000000000E+00;
A(131)=  0.00000000000000E+00; A(132)=  0.16449069541185E-01; A(133)=  0.00000000000000E+00; A(134)= -0.33786380147021E-02; A(135)=  0.14934043240621E+01;
A(136)=  0.10656615979920E+02; A(137)=  0.30000000000000E+01
      Am = Zm + aNm;  Ac = Zc + aNc
     v(1) = Zm/Am; v(2) = aNm/Am; v(3) = (aNm-Zm)/Am
     v(4) = Zc/Zm; v(5) = aNc/aNm; v(6) = (aNc-Zc)/Ac
          v(7) = Ek/Qt
          v(8) =Zc*(Zm-Zc)/(Zm*Zc)
     v(9)  = 0.d0;  if( int(Am/2)*2.ne.Am)     v(9) =1.D0 !odd
     v(10) = 0.d0;  if( int(Zm/2)*2.ne.Zm)     v(10)=1.D0 !odd
     v(11) = 0.d0;  if( int(aNm/2)*2.ne.aNm)   v(11)=1.D0 !odd
```

```fortran
!       print '(8e12.3,3f3.0)',(v(i),i=1,11); pause
      Np1 = 8;   Nc = 3;    Np2 = 15*Np1;    Ns = Np2 + 5; N = Np2 + 5 + 3*Nc +3 !=137
   Amu = 0.d0; do i = 1,Np1; Amu = Amu + a(i)*v(i) + a(i+Np1)*v(i)**2 + a(i+2*Np1)*v(i)**3; enddo  ! 1
   Amu = Amu + a(Np2+1)
       AmCorrection = a(Ns+1)*v(9) + a(Ns+2)*v(10) + a(Ns+3)*v(11)                  ! 1c
   Amc  = dexp( -(Amu + AmCorrection)**2)

   Bmu = 0.d0; do i = 1,Np1; Bmu = Bmu + a(i+3*Np1)*v(i) + a(i+4*Np1)*v(i)**2 + a(i+5*Np1)*v(i)**3; enddo ! 2
      Bmu = Bmu + a(Np2+2)
           BeCorrection = a(Ns+4)*v(9) + a(Ns+5)*v(10) + a(Ns+6)*v(11)              ! 2c
    Be = dexp( Bmu  + BeCorrection)
   Cmu = 0.d0; do i = 1,Np1; Cmu = Cmu + a(i+6*Np1)*v(i) + a(i+7*Np1)*v(i)**2 + a(i+8*Np1)*v(i)**3; enddo   ! 3
      Cmu = Cmu + a(Np2+3)
            CeCorrection = a(Ns+7)*v(9) + a(Ns+8)*v(10) + a(Ns+9)*v(11)                 ! 3c
    Ce = dexp( Cmu  + CeCorrection )
     Pow=1.d0/a(n)
!    print *,Pow,Am,Ac
   R= (Be*((Am - Ac)**Pow + Ac**Pow) - 1.d0)*Ce
   CmnZ = 0.d0; do i = 1,Np1; CmnZ = CmnZ + a(i+9*Np1)*v(i)+ a(i+10*Np1)*v(i)**2 + a(i+11*Np1)*v(i)**3; enddo   ! 4
      CmnZ = CmnZ + a(Np2+4)
    CmnN = 0.d0; do i = 1,Np1; CmnN = CmnN + a(i+12*Np1)*v(i)+ a(i+13*Np1)*v(i)**2 + a(i+14*Np1)*v(i)**3; enddo   ! 5
         CmnN = CmnN + a(Np2+5)
    Pot= Zc*(Zm - Zc)
       PotR= Pot/R
   Zmn = FunZm(Zm,WZ,9)
      aNmn = FunNm(aNm,WN,10)
   Zmn2 = (Zmn-Zm)**2; Wz2 = Wz**2
      aNmn2 = (aNmn-aNm)**2; Wn2 = Wn**2
   CorMagicNumbers = dexp(CmnZ)*dexp(-Zmn2/Wz2)/(Zmn2+Wz2) + dexp(CmnN)*dexp(-aNmn2/Wn2)/(aNmn2+Wn2)
!    print *,r,pot,potr; pause
     Amu = a(n-1)*(1.d0+Amc)
      Te1 = -dexp( a(n-2))*R  + CorMagicNumbers
    Te2 = dsqrt(Am/Amu)*(PotR  - Ek)/dsqrt(Ek)
!    print '(10e14.4)', Amc, Be, Ce, CmnZ, CmnN, R, PotR, Amu, Te1, Te2

   HalfTime =  10.d0**( Te1 + Te2 )    !; print *,halftime; pause
         RETURN; END
!*****************************************************************************C
     Function  FunZm(x,WZ,MnZ)
     IMPLICIT DOUBLE PRECISION (A-H,O-Z)
!    Proton Magic numbers 2,8,14,20,28,50,82,108,124
!    common/nhelp/lexpt,iSP,nPow,nBWp,,MnN,N0,N1,Nqn
     dimension aMn(9) ,aB(9)
```

```fortran
      data aMn/2.d0,8.d0,14.d0,20.d0,28.d0,50.d0,82.d0,108.d0,124.d0/
      data aB/1.d0,5.d0,11.d0,17.d0,24.d0,39.d0,66.d0,95.d0,116.d0/
!     do i = 1, MnZ; aMn(i) =  (a(Nqn+i));  aB(i) =   (a(Nqn + MnZ + i))
!     enddo
      do i = 1,MnZ - 1
       if(x.ge.aB(i).and.x.lt.aB(i+1))      then
        FunZm = int(aMn(i))
         Wz   = (aMn(i+1) - aMn(i))/2
        return; endif
       enddo
       if(x.ge.aB(MnZ))       FunZm = int(aMn(MnZ))
        if(x.ge.aB(MnZ))    WZ  = (aMn(MnZ) - aMn(MnZ-1))/2
!          Print *,'N0=',N0,'N1=',N1,'Nqn=',Nqn, ' N =', N;   pause
      return; end
!*****************************************************************************C
      Function  FunNm(x,WN,MnN)
      IMPLICIT DOUBLE PRECISION (A-H,O-Z)
!    Neutron Magic numbers 2,8,14,20,28,50,82,124,152,202
!    common/nhelp/lexpt,iSP,nPow,nBWp,MnZ,MnN,N0,N1,Nqn
      dimension aMn(10),aB(10)
      data aMn/2.d0,8.d0,14.d0,20.d0,28.d0,50.d0,82.d0,124.d0,152.d0,202.d0/
      data aB/1.d0,5.d0,11.d0,17.d0,24.d0,39.d0,66.d0,103.d0,138.d0,178.d0/
!     do i = 1, MnN; aMn(i) =  (a(Nqn + 2*MnZ + i));  aB(i) =   (a(Nqn + 2*MnZ + MnN + i));
!      enddo
      do i=1,MnN-1
       if(x.ge.aB(i).and.x.lt.aB(i+1))     Then
        FunNm = int(aMn(i))
         WN   = (aMn(i+1) - aMn(i))/2
        return; endif
       enddo
       if(x.ge.aB(MnN))      FunNm = int(aMn(MnN))
        if(x.ge.aB(MnN))     WN   = (aMn(MnN) - aMn(MnN-1))/2
!          Print *,'N0=',N0,'N1=',N1,'Nqn=',Nqn, ' N =', N; pause
      return; end
!*****************************************************************************C
```

# Application B

| Мode | El | A | Z | N | Ac | 3c | Ek | Qt | HtTh | HtExpt | Def | Res | ResLog | Pot | PotR | W | U | R | Amu | CorrMN |
|---|---|---|---|---|---|---|---|---|---|---|---|---|---|---|---|---|---|---|---|---|
| al-dec | Te | 106 | 52 | 54 | 4 | 2 | 4.128 | 4.293 | 0.214629E-11 | 0.190129E-11 | -0.25E-12 | -0.13E+00 | -0.526E-01 | 0.100E+03 | 0.150E+02 | -0.270E+02 | 0.153E+02 | 0.667E+01 | 0.107E+02 | 0.271E+01 |
| al-dec | Te | 108 | 52 | 56 | 4 | 2 | 3.318 | 3.445 | 0.686148E-07 | 0.665450E-07 | -0.21E-08 | -0.31E-01 | -0.133E-01 | 0.100E+03 | 0.151E+02 | -0.237E+02 | 0.165E+02 | 0.663E+01 | 0.107E+02 | 0.583E+01 |
| al-dec | Xe | 112 | 54 | 58 | 4 | 2 | 3.211 | 3.330 | 0.597321E-07 | 0.855578E-07 | 0.26E-07 | 0.29E+00 | 0.156E+00 | 0.104E+03 | 0.155E+02 | -0.254E+02 | 0.182E+02 | 0.669E+01 | 0.107E+02 | 0.435E+01 |
| al-dec | Ba | 114 | 56 | 58 | 4 | 2 | 3.410 | 3.600 | 0.794910E-08 | 0.136259E-07 | 0.57E-08 | 0.31E+00 | 0.234E+00 | 0.108E+03 | 0.158E+02 | -0.286E+02 | 0.205E+02 | 0.683E+01 | 0.107E+02 | 0.184E+01 |
| al-dec | Dy | 150 | 66 | 84 | 4 | 2 | 4.233 | 4.351 | 0.348590E-05 | 0.136322E-04 | 0.10E-04 | 0.74E+00 | 0.592E+00 | 0.128E+03 | 0.182E+02 | -0.237E+02 | 0.182E+02 | 0.705E+01 | 0.107E+02 | 0.771E+01 |
| al-dec | Er | 152 | 68 | 84 | 4 | 2 | 4.804 | 4.934 | 0.232451E-06 | 0.326387E-06 | 0.94E-07 | 0.29E+00 | 0.147E+00 | 0.132E+03 | 0.187E+02 | -0.236E+02 | 0.169E+02 | 0.707E+01 | 0.107E+02 | 0.790E+01 |
| al-dec | Er | 154 | 68 | 86 | 4 | 2 | 4.168 | 4.280 | 0.684485E-05 | 0.703476E-05 | 0.19E-06 | 0.27E-01 | 0.119E-01 | 0.132E+03 | 0.186E+02 | -0.244E+02 | 0.193E+02 | 0.709E+01 | 0.107E+02 | 0.714E+01 |
| al-dec | Yb | 154 | 70 | 84 | 4 | 2 | 5.331 | 5.474 | 0.146108E-07 | 0.129604E-07 | -0.17E-08 | -0.12E+00 | -0.521E-01 | 0.136E+03 | 0.192E+02 | -0.239E+02 | 0.161E+02 | 0.710E+01 | 0.107E+02 | 0.768E+01 |
| al-dec | Yb | 156 | 70 | 86 | 4 | 2 | 4.687 | 4.810 | 0.832414E-06 | 0.823890E-06 | -0.85E-08 | -0.10E-01 | -0.447E-02 | 0.136E+03 | 0.191E+02 | -0.241E+02 | 0.181E+02 | 0.711E+01 | 0.107E+02 | 0.753E+01 |
| al-dec | Hf | 156 | 72 | 84 | 4 | 2 | 5.873 | 6.028 | 0.532475E-09 | 0.728826E-09 | 0.20E-09 | 0.27E+00 | 0.136E+00 | 0.140E+03 | 0.196E+02 | -0.248E+02 | 0.155E+02 | 0.714E+01 | 0.107E+02 | 0.701E+01 |
| al-dec | Hf | 158 | 72 | 86 | 4 | 2 | 5.268 | 5.405 | 0.473018E-07 | 0.903111E-07 | 0.43E-07 | 0.46E+00 | 0.281E+00 | 0.140E+03 | 0.196E+02 | -0.243E+02 | 0.170E+02 | 0.715E+01 | 0.107E+02 | 0.749E+01 |
| al-dec | Hf | 160 | 72 | 88 | 4 | 2 | 4.780 | 4.903 | 0.994156E-06 | 0.430958E-06 | -0.56E-06 | -0.13E+01 | -0.363E+00 | 0.140E+03 | 0.196E+02 | -0.245E+02 | 0.185E+02 | 0.716E+01 | 0.107E+02 | 0.733E+01 |
| al-dec | W | 158 | 74 | 84 | 4 | 2 | 6.445 | 6.613 | 0.156454E-10 | 0.453140E-11 | -0.11E-10 | -0.99E+00 | -0.538E+00 | 0.144E+03 | 0.200E+02 | -0.260E+02 | 0.152E+02 | 0.719E+01 | 0.107E+02 | 0.595E+01 |
| al-dec | W | 160 | 74 | 86 | 4 | 2 | 5.912 | 6.064 | 0.120791E-08 | 0.288362E-08 | 0.17E-08 | 0.58E+00 | 0.378E+00 | 0.144E+03 | 0.200E+02 | -0.251E+02 | 0.161E+02 | 0.719E+01 | 0.107E+02 | 0.693E+01 |
| al-dec | W | 162 | 74 | 88 | 4 | 2 | 5.536 | 5.677 | 0.244961E-07 | 0.430958E-07 | 0.19E-07 | 0.41E+00 | 0.245E+00 | 0.144E+03 | 0.200E+02 | -0.246E+02 | 0.170E+02 | 0.719E+01 | 0.107E+02 | 0.741E+01 |
| al-dec | W | 166 | 74 | 92 | 4 | 2 | 4.739 | 4.856 | 0.210588E-05 | 0.608411E-06 | -0.15E-05 | -0.25E+01 | -0.539E+00 | 0.144E+03 | 0.199E+02 | -0.253E+02 | 0.197E+02 | 0.722E+01 | 0.107E+02 | 0.680E+01 |
| al-dec | Os | 162 | 76 | 86 | 4 | 2 | 6.602 | 6.779 | 0.235763E-10 | 0.649606E-10 | 0.41E-10 | 0.61E+00 | 0.440E+00 | 0.148E+03 | 0.205E+02 | -0.263E+02 | 0.157E+02 | 0.723E+01 | 0.107E+02 | 0.591E+01 |
| al-dec | Os | 166 | 76 | 90 | 4 | 2 | 5.993 | 6.131 | 0.513856E-08 | 0.681294E-08 | 0.17E-08 | 0.25E+00 | 0.122E+00 | 0.148E+03 | 0.204E+02 | -0.248E+02 | 0.165E+02 | 0.724E+01 | 0.107E+02 | 0.742E+01 |
| al-dec | Os | 170 | 76 | 94 | 4 | 2 | 5.407 | 5.539 | 0.156456E-06 | 0.233541E-06 | 0.77E-07 | 0.32E+00 | 0.174E+00 | 0.148E+03 | 0.204E+02 | -0.251E+02 | 0.183E+02 | 0.726E+01 | 0.107E+02 | 0.721E+01 |
| al-dec | Os | 174 | 76 | 98 | 4 | 2 | 4.760 | 4.872 | 0.204551E-05 | 0.139428E-05 | -0.65E-06 | -0.47E+00 | -0.166E+00 | 0.148E+03 | 0.203E+02 | -0.269E+02 | 0.212E+02 | 0.730E+01 | 0.107E+02 | 0.560E+01 |
| al-dec | Pt | 168 | 78 | 90 | 4 | 2 | 6.832 | 6.990 | 0.562847E-10 | 0.640099E-10 | 0.77E-11 | 0.12E+00 | 0.559E-01 | 0.152E+03 | 0.209E+02 | -0.257E+02 | 0.154E+02 | 0.728E+01 | 0.107E+02 | 0.674E+01 |
| al-dec | Pt | 170 | 78 | 92 | 4 | 2 | 6.548 | 6.704 | 0.720274E-09 | 0.441415E-09 | -0.28E-09 | -0.62E+00 | -0.213E+00 | 0.152E+03 | 0.209E+02 | -0.250E+02 | 0.159E+02 | 0.728E+01 | 0.107E+02 | 0.741E+01 |
| al-dec | Pt | 174 | 78 | 96 | 4 | 2 | 6.039 | 6.184 | 0.268500E-07 | 0.277588E-07 | 0.91E-09 | 0.27E-01 | 0.145E-01 | 0.152E+03 | 0.208E+02 | -0.248E+02 | 0.173E+02 | 0.730E+01 | 0.107E+02 | 0.767E+01 |
| al-dec | Pt | 176 | 78 | 98 | 4 | 2 | 5.753 | 5.886 | 0.965279E-07 | 0.199635E-06 | 0.10E-06 | 0.51E+00 | 0.316E+00 | 0.152E+03 | 0.208E+02 | -0.253E+02 | 0.183E+02 | 0.732E+01 | 0.107E+02 | 0.724E+01 |
| al-dec | Pt | 178 | 78 | 100 | 4 | 2 | 5.446 | 5.573 | 0.292908E-06 | 0.668619E-06 | 0.38E-06 | 0.56E+00 | 0.358E+00 | 0.152E+03 | 0.207E+02 | -0.262E+02 | 0.196E+02 | 0.734E+01 | 0.107E+02 | 0.648E+01 |
| al-dec | Pt | 180 | 78 | 102 | 4 | 2 | 5.140 | 5.275 | 0.856994E-06 | 0.177453E-05 | 0.92E-06 | 0.52E+00 | 0.316E+00 | 0.152E+03 | 0.207E+02 | -0.273E+02 | 0.212E+02 | 0.735E+01 | 0.107E+02 | 0.548E+01 |
| al-dec | Pt | 182 | 78 | 104 | 4 | 2 | 4.843 | 4.951 | 0.324253E-05 | 0.507643E-05 | 0.18E-05 | 0.35E+00 | 0.195E+00 | 0.152E+03 | 0.206E+02 | -0.287E+02 | 0.232E+02 | 0.737E+01 | 0.107E+02 | 0.417E+01 |
| al-dec | Hg | 174 | 80 | 94 | 4 | 2 | 7.067 | 7.233 | 0.971547E-10 | 0.633762E-10 | -0.34E-10 | -0.53E+00 | -0.186E+00 | 0.156E+03 | 0.213E+02 | -0.254E+02 | 0.154E+02 | 0.733E+01 | 0.107E+02 | 0.722E+01 |
| al-dec | Hg | 176 | 80 | 96 | 4 | 2 | 6.751 | 6.897 | 0.137935E-08 | 0.684463E-09 | -0.69E-09 | -0.10E+01 | -0.304E+00 | 0.156E+03 | 0.213E+02 | -0.249E+02 | 0.161E+02 | 0.733E+01 | 0.107E+02 | 0.774E+01 |
| al-dec | Hg | 180 | 80 | 100 | 4 | 2 | 6.119 | 6.258 | 0.978285E-07 | 0.817553E-07 | -0.16E-07 | -0.20E+00 | -0.779E-01 | 0.156E+03 | 0.212E+02 | -0.250E+02 | 0.180E+02 | 0.736E+01 | 0.107E+02 | 0.776E+01 |
| al-dec | Hg | 182 | 80 | 102 | 4 | 2 | 5.867 | 5.999 | 0.277569E-06 | 0.343182E-06 | 0.66E-07 | 0.19E+00 | 0.922E-01 | 0.156E+03 | 0.212E+02 | -0.256E+02 | 0.191E+02 | 0.738E+01 | 0.107E+02 | 0.719E+01 |
| al-dec | Hg | 184 | 80 | 104 | 4 | 2 | 5.535 | 5.662 | 0.680309E-06 | 0.979162E-06 | 0.30E-06 | 0.30E+00 | 0.158E+00 | 0.156E+03 | 0.211E+02 | -0.268E+02 | 0.206E+02 | 0.739E+01 | 0.107E+02 | 0.611E+01 |
| al-dec | Pb | 186 | 82 | 104 | 4 | 2 | 6.331 | 6.470 | 0.479043E-07 | 0.152737E-06 | 0.10E-06 | 0.68E+00 | 0.504E+00 | 0.160E+03 | 0.216E+02 | -0.258E+02 | 0.185E+02 | 0.742E+01 | 0.107E+02 | 0.723E+01 |
| al-dec | Pb | 188 | 82 | 106 | 4 | 2 | 5.983 | 6.109 | 0.523744E-06 | 0.795371E-06 | 0.27E-06 | 0.34E+00 | 0.181E+00 | 0.160E+03 | 0.215E+02 | -0.263E+02 | 0.200E+02 | 0.743E+01 | 0.107E+02 | 0.684E+01 |
| al-dec | Pb | 190 | 82 | 108 | 4 | 2 | 5.581 | 5.870 | 0.614460E-05 | 0.224985E-05 | -0.39E-05 | -0.17E+01 | -0.436E+00 | 0.160E+03 | 0.213E+02 | -0.267E+02 | 0.215E+02 | 0.750E+01 | 0.107E+02 | 0.668E+01 |
| al-dec | Po | 188 | 84 | 104 | 4 | 2 | 7.911 | 8.082 | 0.931941E-11 | 0.871422E-11 | -0.61E-12 | -0.69E-01 | -0.292E-01 | 0.164E+03 | 0.220E+02 | -0.261E+02 | 0.151E+02 | 0.744E+01 | 0.107E+02 | 0.700E+01 |
| al-dec | Po | 190 | 84 | 106 | 4 | 2 | 7.537 | 7.693 | 0.159447E-09 | 0.779527E-10 | -0.81E-10 | -0.10E+01 | -0.311E+00 | 0.164E+03 | 0.220E+02 | -0.260E+02 | 0.162E+02 | 0.745E+01 | 0.107E+02 | 0.721E+01 |
| al-dec | Po | 192 | 84 | 108 | 4 | 2 | 7.167 | 7.320 | 0.170866E-08 | 0.105204E-08 | -0.66E-09 | -0.12E+00 | -0.211E+00 | 0.164E+03 | 0.220E+02 | -0.262E+02 | 0.174E+02 | 0.747E+01 | 0.107E+02 | 0.705E+01 |
| al-dec | Po | 194 | 84 | 110 | 4 | 2 | 6.843 | 6.987 | 0.113332E-07 | 0.124217E-07 | 0.11E-08 | 0.79E-01 | 0.398E-01 | 0.164E+03 | 0.219E+02 | -0.268E+02 | 0.188E+02 | 0.749E+01 | 0.107E+02 | 0.654E+01 |
| al-dec | Po | 196 | 84 | 112 | 4 | 2 | 6.520 | 6.657 | 0.807782E-07 | 0.181573E-06 | 0.10E-06 | 0.53E+00 | 0.352E+00 | 0.164E+03 | 0.219E+02 | -0.276E+02 | 0.205E+02 | 0.750E+01 | 0.107E+02 | 0.581E+01 |
| al-dec | Po | 198 | 84 | 114 | 4 | 2 | 6.182 | 6.309 | 0.101680E-05 | 0.336527E-05 | 0.23E-05 | 0.69E+00 | 0.520E+00 | 0.164E+03 | 0.218E+02 | -0.285E+02 | 0.225E+02 | 0.752E+01 | 0.107E+02 | 0.496E+01 |
| al-dec | Po | 200 | 84 | 116 | 4 | 2 | 5.862 | 5.981 | 0.219582E-04 | 0.218648E-04 | -0.93E-07 | -0.43E-02 | -0.185E-02 | 0.164E+03 | 0.218E+02 | -0.295E+02 | 0.248E+02 | 0.754E+01 | 0.107E+02 | 0.410E+01 |
| al-dec | Po | 202 | 84 | 118 | 4 | 2 | 5.588 | 5.701 | 0.586716E-03 | 0.847973E-04 | -0.50E-03 | -0.59E+01 | -0.840E+00 | 0.164E+03 | 0.217E+02 | -0.304E+02 | 0.271E+02 | 0.755E+01 | 0.107E+02 | 0.327E+01 |
| al-dec | Po | 206 | 84 | 122 | 4 | 2 | 5.224 | 5.327 | 0.870802E-01 | 0.240931E-01 | -0.63E-01 | -0.26E+01 | -0.558E+00 | 0.164E+03 | 0.216E+02 | -0.319E+02 | 0.308E+02 | 0.758E+01 | 0.107E+02 | 0.186E+01 |
| al-dec | Po | 212 | 84 | 128 | 4 | 2 | 8.785 | 8.954 | 0.866765E-14 | 0.947474E-14 | 0.81E-15 | 0.80E-01 | 0.387E-01 | 0.164E+03 | 0.216E+02 | -0.333E+02 | 0.192E+02 | 0.760E+01 | 0.107E+02 | 0.529E+00 |

```
al-dec Po     214 84 130   4 2   7.687    7.833  0.831166E-11  0.520635E-11  -0.31E-11  -0.60E+00  -0.203E+00   0.164E+03   0.216E+02  -0.335E+02   0.225E+02   0.760E+01   0.107E+02   0.307E+00
al-dec Po     216 84 132   4 2   6.778    6.906  0.889995E-08  0.459477E-08  -0.43E-08  -0.82E+00  -0.287E+00   0.164E+03   0.216E+02  -0.337E+02   0.256E+02   0.760E+01   0.107E+02   0.168E+00
al-dec Po     218 84 134   4 2   6.002    6.115  0.124952E-04  0.589018E-05  -0.66E-05  -0.11E+01  -0.327E+00   0.164E+03   0.216E+02  -0.337E+02   0.288E+02   0.759E+01   0.107E+02   0.872E-01
al-dec Rn     196 86 110   4 2   7.465    7.617  0.492901E-09  0.148934E-09  -0.34E-09  -0.23E+01  -0.520E+00   0.168E+03   0.224E+02  -0.267E+02   0.173E+02   0.751E+01   0.107E+02   0.678E+01
al-dec Rn     198 86 112   4 2   7.205    7.349  0.362751E-08  0.205973E-08  -0.16E-08  -0.76E+00  -0.246E+00   0.168E+03   0.223E+02  -0.270E+02   0.185E+02   0.753E+01   0.107E+02   0.654E+01
al-dec Rn     200 86 114   4 2   6.902    7.043  0.307477E-07  0.304206E-07  -0.33E-09  -0.10E-01  -0.465E-02   0.168E+03   0.223E+02  -0.275E+02   0.200E+02   0.754E+01   0.107E+02   0.603E+01
al-dec Rn     202 86 116   4 2   6.639    6.774  0.240088E-06  0.307374E-06   0.67E-07   0.22E+00   0.107E+00   0.168E+03   0.222E+02  -0.283E+02   0.217E+02   0.756E+01   0.107E+02   0.533E+01
al-dec Rn     206 86 120   4 2   6.260    6.384  0.109894E-04  0.107803E-04  -0.21E-06  -0.19E-01  -0.834E-02   0.168E+03   0.221E+02  -0.302E+02   0.252E+02   0.759E+01   0.107E+02   0.364E+01
al-dec Rn     208 86 122   4 2   6.140    6.261  0.475207E-04  0.462963E-04  -0.12E-05  -0.26E-01  -0.113E-01   0.168E+03   0.221E+02  -0.311E+02   0.267E+02   0.760E+01   0.107E+02   0.278E+01
al-dec Rn     210 86 124   4 2   6.041    6.159  0.134776E-03  0.273785E-03   0.14E-03   0.51E+00   0.308E+00   0.168E+03   0.221E+02  -0.319E+02   0.280E+02   0.762E+01   0.107E+02   0.200E+01
al-dec Rn     212 86 126   4 2   6.264    6.385  0.132972E-04  0.454407E-04   0.32E-04   0.70E+00   0.534E+00   0.168E+03   0.220E+02  -0.326E+02   0.277E+02   0.763E+01   0.107E+02   0.134E+01
al-dec Rn     218 86 132   4 2   7.129    7.263  0.257435E-08  0.110908E-08  -0.15E-08  -0.12E+01  -0.366E+00   0.168E+03   0.220E+02  -0.338E+02   0.252E+02   0.764E+01   0.107E+02   0.272E+00
al-dec Rn     220 86 134   4 2   6.288    6.405  0.373445E-05  0.176186E-05  -0.20E-05  -0.11E+01  -0.326E+00   0.168E+03   0.220E+02  -0.339E+02   0.284E+02   0.764E+01   0.107E+02   0.142E+00
al-dec Rn     222 86 136   4 2   5.489    5.590  0.173893E-01  0.104682E-01  -0.69E-02  -0.66E+00  -0.220E+00   0.168E+03   0.220E+02  -0.339E+02   0.322E+02   0.764E+01   0.107E+02   0.705E-01
al-dec Ra     204 88 116   4 2   7.486    7.636  0.215894E-08  0.180622E-08  -0.35E-09  -0.20E+00  -0.775E-01   0.172E+03   0.227E+02  -0.277E+02   0.191E+02   0.758E+01   0.107E+02   0.600E+01
al-dec Ra     206 88 118   4 2   7.269    7.415  0.123206E-07  0.760514E-08  -0.47E-08  -0.57E+00  -0.210E+00   0.172E+03   0.226E+02  -0.284E+02   0.205E+02   0.759E+01   0.107E+02   0.544E+01
al-dec Ra     208 88 120   4 2   7.133    7.273  0.404439E-07  0.411945E-07   0.75E-09   0.18E-01   0.799E-02   0.172E+03   0.226E+02  -0.292E+02   0.218E+02   0.761E+01   0.107E+02   0.467E+01
al-dec Ra     210 88 122   4 2   7.016    7.157  0.109013E-06  0.117246E-06   0.82E-08   0.70E-01   0.316E-01   0.172E+03   0.226E+02  -0.302E+02   0.232E+02   0.763E+01   0.107E+02   0.375E+01
al-dec Ra     212 88 124   4 2   6.899    7.032  0.299349E-06  0.411945E-06   0.11E-06   0.27E+00   0.139E+00   0.172E+03   0.225E+02  -0.312E+02   0.247E+02   0.764E+01   0.107E+02   0.282E+01
al-dec Ra     214 88 126   4 2   7.137    7.272  0.433270E-07  0.779527E-07   0.35E-07   0.44E+00   0.255E+00   0.172E+03   0.225E+02  -0.321E+02   0.247E+02   0.765E+01   0.107E+02   0.197E+01
al-dec Ra     220 88 132   4 2   7.453    7.592  0.122543E-08  0.567217E-09  -0.66E-09  -0.12E+01  -0.335E+00   0.172E+03   0.224E+02  -0.337E+02   0.248E+02   0.768E+01   0.107E+02   0.438E+00
al-dec Ra     222 88 134   4 2   6.559    6.679  0.173255E-05  0.106472E-05  -0.67E-06  -0.63E+00  -0.211E+00   0.172E+03   0.224E+02  -0.340E+02   0.282E+02   0.768E+01   0.107E+02   0.234E+00
al-dec Ra     224 88 136   4 2   5.685    5.789  0.108302E-01  0.100205E-01  -0.81E-03  -0.80E-01  -0.337E-01   0.172E+03   0.224E+02  -0.341E+02   0.321E+02   0.768E+01   0.107E+02   0.119E+00
al-dec Ra     226 88 138   4 2   4.784    4.871  0.890150E+03  0.160000E+04   0.71E+03   0.44E+00   0.255E+00   0.172E+03   0.224E+02  -0.341E+02   0.371E+02   0.768E+01   0.107E+02   0.553E-01
al-dec Th     210 90 120   4 2   7.899    8.069  0.512030E-09  0.507009E-09  -0.50E-11  -0.99E-02  -0.428E-02   0.176E+03   0.231E+02  -0.286E+02   0.193E+02   0.763E+01   0.107E+02   0.534E+01
al-dec Th     216 90 126   4 2   7.922    8.072  0.358089E-09  0.823890E-09   0.47E-09   0.57E+00   0.362E+00   0.176E+03   0.229E+02  -0.315E+02   0.221E+02   0.768E+01   0.107E+02   0.263E+01
al-dec Th     222 90 132   4 2   7.980    8.127  0.129385E-09  0.649606E-10  -0.64E-10  -0.99E+00  -0.299E+00   0.176E+03   0.228E+02  -0.337E+02   0.238E+02   0.771E+01   0.107E+02   0.664E+00
al-dec Th     224 90 134   4 2   7.170    7.298  0.515828E-07  0.332725E-07  -0.18E-07  -0.54E+00  -0.190E+00   0.176E+03   0.228E+02  -0.340E+02   0.267E+02   0.772E+01   0.107E+02   0.370E+00
al-dec Th     226 90 136   4 2   6.337    6.451  0.816109E-04  0.583695E-04  -0.23E-04  -0.40E+00  -0.146E+00   0.176E+03   0.228E+02  -0.342E+02   0.301E+02   0.772E+01   0.107E+02   0.195E+00
al-dec Th     228 90 138   4 2   5.423    5.520  0.160891E+01  0.191200E+01   0.30E+00   0.16E+00   0.750E-01   0.176E+03   0.228E+02  -0.343E+02   0.345E+02   0.772E+01   0.107E+02   0.953E-01
al-dec Th     230 90 140   4 2   4.687    4.770  0.351278E+05  0.754000E+05   0.40E+05   0.53E+00   0.332E+00   0.176E+03   0.228E+02  -0.343E+02   0.389E+02   0.772E+01   0.107E+02   0.664E-01
al-dec U      226 92 134   4 2   7.570    7.701  0.858854E-08  0.852410E-08  -0.64E-10  -0.76E-02  -0.327E-02   0.180E+03   0.232E+02  -0.339E+02   0.259E+02   0.775E+01   0.107E+02   0.539E+00
al-dec U      228 92 136   4 2   6.680    6.803  0.174777E-04  0.173017E-04  -0.18E-06  -0.10E-01  -0.440E-02   0.180E+03   0.232E+02  -0.342E+02   0.295E+02   0.775E+01   0.107E+02   0.298E+00
al-dec U      230 92 138   4 2   5.888    5.993  0.516432E-01  0.553867E-01   0.37E-02   0.68E-01   0.304E-01   0.180E+03   0.232E+02  -0.344E+02   0.331E+02   0.776E+01   0.107E+02   0.154E+00
al-dec U      232 92 140   4 2   5.320    5.414  0.495098E+02  0.689000E+02   0.19E+02   0.28E+00   0.144E+00   0.180E+03   0.232E+02  -0.344E+02   0.361E+02   0.776E+01   0.107E+02   0.107E+00
al-dec U      234 92 142   4 2   4.775    4.858  0.102840E+06  0.245500E+06   0.14E+06   0.57E+00   0.378E+00   0.180E+03   0.232E+02  -0.345E+02   0.395E+02   0.776E+01   0.107E+02   0.764E-01
al-dec U      236 92 144   4 2   4.494    4.573  0.108986E+08  0.234200E+08   0.13E+08   0.53E+00   0.332E+00   0.180E+03   0.232E+02  -0.345E+02   0.415E+02   0.776E+01   0.107E+02   0.558E-01
al-dec U      238 92 146   4 2   4.198    4.270  0.243387E+10  0.446800E+10   0.20E+10   0.45E+00   0.264E+00   0.180E+03   0.232E+02  -0.345E+02   0.439E+02   0.775E+01   0.107E+02   0.415E-01
al-dec Pu     232 94 138   4 2   6.600    6.716  0.175929E-03  0.648338E-04  -0.11E-03  -0.17E+01  -0.434E+00   0.184E+03   0.236E+02  -0.345E+02   0.307E+02   0.779E+01   0.107E+02   0.224E+00
al-dec Pu     236 94 142   4 2   5.768    5.867  0.199322E+01  0.285800E+01   0.86E+00   0.29E+00   0.157E+00   0.184E+03   0.236E+02  -0.346E+02   0.349E+02   0.780E+01   0.107E+02   0.116E+00
al-dec Pu     238 94 144   4 2   5.499    5.593  0.633980E+02  0.877000E+02   0.24E+02   0.28E+00   0.141E+00   0.184E+03   0.236E+02  -0.346E+02   0.364E+02   0.780E+01   0.107E+02   0.851E-01
al-dec Pu     240 94 146   4 2   5.168    5.256  0.618270E+04  0.656100E+04   0.38E+03   0.58E-01   0.258E-01   0.184E+03   0.236E+02  -0.347E+02   0.385E+02   0.780E+01   0.107E+02   0.628E-01
al-dec Pu     242 94 148   4 2   4.902    4.984  0.384427E+06  0.375000E+06  -0.94E+04  -0.24E-01  -0.108E-01   0.184E+03   0.236E+02  -0.347E+02   0.402E+02   0.780E+01   0.107E+02   0.465E-01
al-dec Pu     244 94 150   4 2   4.589    4.665  0.748901E+08  0.800000E+08   0.51E+07   0.64E-01   0.287E-01   0.184E+03   0.236E+02  -0.346E+02   0.425E+02   0.779E+01   0.107E+02   0.344E-01
al-dec Cm     240 96 144   4 2   6.290    6.398  0.476835E-01  0.739220E-01   0.26E-01   0.35E+00   0.190E+00   0.188E+03   0.240E+02  -0.348E+02   0.335E+02   0.784E+01   0.107E+02   0.110E+00
al-dec Cm     242 96 146   4 2   6.113    6.216  0.414466E+00  0.446105E+00   0.32E-01   0.71E-01   0.319E-01   0.188E+03   0.240E+02  -0.348E+02   0.344E+02   0.784E+01   0.107E+02   0.848E-01
al-dec Cm     244 96 148   4 2   5.805    5.902  0.182116E+02  0.181100E+02  -0.10E+00  -0.56E-02  -0.243E-02   0.188E+03   0.240E+02  -0.348E+02   0.361E+02   0.784E+01   0.107E+02   0.646E-01
al-dec Cm     246 96 150   4 2   5.386    5.475  0.461486E+04  0.476000E+04   0.15E+03   0.30E-01   0.134E-01   0.188E+03   0.240E+02  -0.348E+02   0.385E+02   0.784E+01   0.107E+02   0.486E-01
al-dec Cm     248 96 152   4 2   5.078    5.162  0.483099E+06  0.348000E+06  -0.14E+06  -0.38E+00  -0.142E+00   0.188E+03   0.240E+02  -0.348E+02   0.405E+02   0.783E+01   0.107E+02   0.359E-01
```

| | | | | | | | | | | | | | | | | | |
|---|---|---|---|---|---|---|---|---|---|---|---|---|---|---|---|---|---|
| al-dec | Cf | 244 | 98 | 146 | 4 | 2 | 7.209 | 7.329 | 0.478898E-04 | 0.368849E-04 | -0.11E-04 | -0.30E+00 | -0.113E+00 | 0.192E+03 | 0.244E+02 | -0.349E+02 | 0.306E+02 | 0.787E+01 | 0.107E+02 | 0.169E+00 |
| al-dec | Cf | 246 | 98 | 148 | 4 | 2 | 6.750 | 6.862 | 0.431270E-02 | 0.407255E-02 | -0.24E-03 | -0.59E-01 | -0.249E-01 | 0.192E+03 | 0.244E+02 | -0.349E+02 | 0.326E+02 | 0.788E+01 | 0.107E+02 | 0.121E+00 |
| al-dec | Cf | 248 | 98 | 150 | 4 | 2 | 6.258 | 6.361 | 0.871915E+00 | 0.914442E+00 | 0.43E-01 | 0.47E-01 | 0.207E-01 | 0.192E+03 | 0.244E+02 | -0.350E+02 | 0.349E+02 | 0.788E+01 | 0.107E+02 | 0.853E-01 |
| al-dec | Cf | 250 | 98 | 152 | 4 | 2 | 6.030 | 6.128 | 0.149600E+02 | 0.130800E+02 | -0.19E+01 | -0.14E+00 | -0.583E-01 | 0.192E+03 | 0.244E+02 | -0.350E+02 | 0.362E+02 | 0.788E+01 | 0.107E+02 | 0.599E-01 |
| al-dec | Cf | 252 | 98 | 154 | 4 | 2 | 6.118 | 6.217 | 0.806206E+01 | 0.264500E+01 | -0.54E+01 | -0.20E+01 | -0.484E+00 | 0.192E+03 | 0.244E+02 | -0.350E+02 | 0.359E+02 | 0.787E+01 | 0.107E+02 | 0.418E-01 |
| al-dec | Fm | 246 | 100 | 146 | 4 | 2 | 8.237 | 8.377 | 0.630358E-07 | 0.487997E-07 | -0.14E-07 | -0.28E+00 | -0.111E+00 | 0.196E+03 | 0.248E+02 | -0.347E+02 | 0.275E+02 | 0.790E+01 | 0.107E+02 | 0.456E+00 |
| al-dec | Fm | 248 | 100 | 148 | 4 | 2 | 7.870 | 7.994 | 0.124411E-05 | 0.111225E-05 | -0.13E-06 | -0.12E+00 | -0.487E-01 | 0.196E+03 | 0.248E+02 | -0.349E+02 | 0.290E+02 | 0.791E+01 | 0.107E+02 | 0.309E+00 |
| al-dec | Fm | 250 | 100 | 150 | 4 | 2 | 7.436 | 7.557 | 0.547350E-04 | 0.627424E-04 | 0.80E-05 | 0.13E+00 | 0.593E-01 | 0.196E+03 | 0.248E+02 | -0.350E+02 | 0.308E+02 | 0.791E+01 | 0.107E+02 | 0.204E+00 |
| al-dec | Fm | 252 | 100 | 152 | 4 | 2 | 7.039 | 7.153 | 0.235340E-02 | 0.289642E-02 | 0.54E-03 | 0.19E+00 | 0.902E-01 | 0.196E+03 | 0.248E+02 | -0.351E+02 | 0.325E+02 | 0.792E+01 | 0.107E+02 | 0.134E+00 |
| al-dec | Fm | 254 | 100 | 154 | 4 | 2 | 7.192 | 7.307 | 0.675376E-03 | 0.369610E-03 | -0.31E-03 | -0.82E+00 | -0.262E+00 | 0.196E+03 | 0.248E+02 | -0.352E+02 | 0.320E+02 | 0.792E+01 | 0.107E+02 | 0.864E-01 |
| al-dec | No | 252 | 102 | 150 | 4 | 2 | 8.415 | 8.548 | 0.169029E-06 | 0.776358E-07 | -0.91E-07 | -0.12E+01 | -0.338E+00 | 0.200E+03 | 0.252E+02 | -0.348E+02 | 0.280E+02 | 0.794E+01 | 0.107E+02 | 0.613E+00 |
| al-dec | No | 256 | 102 | 154 | 4 | 2 | 8.448 | 8.477 | 0.415708E-07 | 0.922123E-07 | 0.51E-07 | 0.54E+00 | 0.346E+00 | 0.200E+03 | 0.250E+02 | -0.353E+02 | 0.279E+02 | 0.799E+01 | 0.107E+02 | 0.252E+00 |
| al-dec | Hs | 270 | 108 | 162 | 4 | 2 | 9.120 | 9.300 | 0.303468E-06 | 0.114077E-06 | -0.19E-06 | -0.15E+01 | -0.425E+00 | 0.212E+03 | 0.263E+02 | -0.352E+02 | 0.287E+02 | 0.805E+01 | 0.107E+02 | 0.686E+00 |
| al-dec | Ds | 270 | 110 | 160 | 4 | 2 | 9.990 | 11.200 | 0.329888E-08 | 0.316881E-08 | -0.13E-09 | -0.40E-01 | -0.175E-01 | 0.216E+03 | 0.257E+02 | -0.335E+02 | 0.250E+02 | 0.839E+01 | 0.107E+02 | 0.386E+01 |
| al-dec | Fl | 286 | 114 | 172 | 4 | 2 | 10.190 | 10.330 | 0.606361E-08 | 0.411945E-08 | -0.19E-08 | -0.36E+00 | -0.168E+00 | 0.224E+03 | 0.274E+02 | -0.362E+02 | 0.280E+02 | 0.817E+01 | 0.107E+02 | 0.201E+00 |
| al-dec | Fl | 288 | 114 | 174 | 4 | 2 | 9.950 | 10.140 | 0.415211E-07 | 0.148934E-07 | -0.27E-07 | -0.12E+01 | -0.445E+00 | 0.224E+03 | 0.274E+02 | -0.362E+02 | 0.288E+02 | 0.816E+01 | 0.107E+02 | 0.120E+00 |
| al-dec | Lv | 290 | 116 | 174 | 4 | 2 | 10.840 | 10.990 | 0.293309E-09 | 0.253505E-09 | -0.40E-10 | -0.11E+00 | -0.633E-01 | 0.228E+03 | 0.278E+02 | -0.364E+02 | 0.269E+02 | 0.820E+01 | 0.107E+02 | 0.134E+00 |
| al-dec | Lv | 292 | 116 | 176 | 4 | 2 | 10.660 | 10.774 | 0.865384E-09 | 0.760514E-09 | -0.10E-09 | -0.83E-01 | -0.561E-01 | 0.228E+03 | 0.278E+02 | -0.365E+02 | 0.274E+02 | 0.821E+01 | 0.107E+02 | 0.824E-01 |
| al-dec | Ei | 294 | 118 | 176 | 4 | 2 | 11.650 | 11.810 | 0.112373E-10 | 0.443633E-10 | 0.33E-10 | 0.42E+00 | 0.596E+00 | 0.232E+03 | 0.282E+02 | -0.364E+02 | 0.254E+02 | 0.824E+01 | 0.107E+02 | 0.314E+00 |
| cl-dec | Ra | 224 | 88 | 136 | 14 | 6 | 28.630 | 30.535 | 0.771158E+09 | 0.230000E+09 | -0.54E+09 | -0.23E+01 | -0.525E+00 | 0.492E+03 | 0.824E+02 | -0.265E+02 | 0.354E+02 | 0.597E+01 | 0.107E+02 | 0.619E-01 |
| cl-dec | Ra | 226 | 88 | 138 | 14 | 6 | 26.460 | 28.689 | 0.193499E+14 | 0.530000E+14 | 0.34E+14 | 0.63E+00 | 0.438E+00 | 0.492E+03 | 0.816E+02 | -0.268E+02 | 0.401E+02 | 0.603E+01 | 0.107E+02 | 0.322E-01 |
| cl-dec | Th | 228 | 90 | 138 | 20 | 8 | 40.250 | 44.730 | 0.170595E+14 | 0.168000E+14 | -0.26E+12 | -0.15E-01 | -0.666E-02 | 0.656E+03 | 0.109E+03 | -0.265E+02 | 0.398E+02 | 0.601E+01 | 0.107E+02 | 0.212E+00 |
| cl-dec | Th | 232 | 90 | 142 | 26 | 10 | 49.700 | 55.975 | 0.120785E+23 | 0.120000E+23 | -0.78E+20 | -0.55E-02 | -0.283E-02 | 0.800E+03 | 0.126E+03 | -0.281E+02 | 0.502E+02 | 0.634E+01 | 0.107E+02 | 0.996E-01 |
| cl-dec | U | 234 | 92 | 142 | 24 | 10 | 52.810 | 58.843 | 0.576533E+18 | 0.270000E+19 | 0.21E+19 | 0.66E+00 | 0.671E+00 | 0.820E+03 | 0.130E+03 | -0.281E+02 | 0.458E+02 | 0.633E+01 | 0.107E+02 | 0.946E-01 |
| cl-dec | U | 234 | 92 | 142 | 26 | 10 | 52.870 | 59.476 | 0.103411E+20 | 0.270000E+19 | -0.76E+19 | -0.24E+01 | -0.583E+00 | 0.820E+03 | 0.128E+03 | -0.284E+02 | 0.474E+02 | 0.643E+01 | 0.107E+02 | 0.234E+00 |
| cl-dec | U | 234 | 92 | 142 | 28 | 12 | 65.260 | 74.129 | 0.177191E+19 | 0.170000E+19 | -0.72E+17 | -0.38E-01 | -0.180E-01 | 0.960E+03 | 0.147E+03 | -0.290E+02 | 0.473E+02 | 0.654E+01 | 0.107E+02 | 0.125E+00 |
| cl-dec | Pu | 236 | 94 | 142 | 28 | 12 | 70.220 | 79.669 | 0.149232E+15 | 0.110000E+15 | -0.39E+14 | -0.28E+00 | -0.132E+00 | 0.984E+03 | 0.148E+03 | -0.294E+02 | 0.436E+02 | 0.666E+01 | 0.107E+02 | 0.233E+00 |
| cl-dec | Pu | 238 | 94 | 144 | 28 | 12 | 67.320 | 75.930 | 0.458538E+18 | 0.150000E+19 | 0.10E+19 | 0.60E+00 | 0.515E+00 | 0.984E+03 | 0.149E+03 | -0.293E+02 | 0.470E+02 | 0.661E+01 | 0.107E+02 | 0.107E+00 |
| cl-dec | Pu | 238 | 94 | 144 | 30 | 12 | 67.000 | 77.032 | 0.153951E+19 | 0.150000E+19 | -0.40E+17 | -0.22E-01 | -0.113E-01 | 0.984E+03 | 0.149E+03 | -0.291E+02 | 0.473E+02 | 0.660E+01 | 0.107E+02 | 0.293E+00 |
| cl-dec | Pu | 238 | 94 | 144 | 32 | 14 | 78.950 | 91.209 | 0.647618E+18 | 0.630000E+18 | -0.18E+17 | -0.27E-01 | -0.120E-01 | 0.112E+04 | 0.168E+03 | -0.293E+02 | 0.471E+02 | 0.668E+01 | 0.107E+02 | 0.450E+00 |
| cl-dec | Cm | 242 | 96 | 146 | 34 | 14 | 82.880 | 96.431 | 0.814302E+16 | 0.600000E+16 | -0.21E+16 | -0.30E+00 | -0.133E+00 | 0.115E+04 | 0.170E+03 | -0.298E+02 | 0.457E+02 | 0.674E+01 | 0.107E+02 | 0.190E+00 |
| sp-dec | U | 234 | 92 | 142 | 92 | 38 | 163.000 | 176.100 | 0.891838E+16 | 0.150000E+17 | 0.61E+16 | 0.36E+00 | 0.226E+00 | 0.205E+04 | 0.274E+03 | -0.248E+02 | 0.408E+02 | 0.749E+01 | 0.107E+02 | 0.852E+01 |
| sp-dec | U | 234 | 92 | 142 | 98 | 37 | 164.000 | 185.310 | 0.243662E+17 | 0.160000E+17 | -0.84E+16 | -0.46E+00 | -0.183E+00 | 0.204E+04 | 0.196E+03 | 0.470E+01 | 0.117E+02 | 0.104E+02 | 0.107E+02 | 0.509E+02 |
| sp-dec | U | 236 | 92 | 144 | 94 | 38 | 165.000 | 175.600 | 0.429037E+17 | 0.250000E+17 | -0.18E+17 | -0.69E+00 | -0.235E+00 | 0.205E+04 | 0.286E+03 | -0.277E+02 | 0.443E+02 | 0.718E+01 | 0.107E+02 | 0.431E+01 |
| sp-dec | Pu | 236 | 92 | 144 | 98 | 37 | 165.000 | 181.490 | 0.310904E+17 | 0.200000E+17 | -0.11E+17 | -0.53E+00 | -0.192E+00 | 0.204E+04 | 0.235E+03 | -0.915E+01 | 0.256E+02 | 0.866E+01 | 0.107E+02 | 0.294E+02 |
| sp-dec | Pu | 236 | 94 | 142 | 94 | 38 | 176.300 | 185.300 | 0.186686E+10 | 0.150000E+10 | -0.37E+09 | -0.12E-01 | -0.950E-01 | 0.213E+04 | 0.269E+03 | -0.237E+02 | 0.330E+02 | 0.790E+01 | 0.107E+02 | 0.115E+02 |
| sp-dec | Pu | 238 | 94 | 144 | 99 | 41 | 174.600 | 198.000 | 0.597365E+11 | 0.475000E+11 | -0.12E+11 | -0.25E+00 | -0.995E-01 | 0.217E+04 | 0.276E+03 | -0.254E+02 | 0.362E+02 | 0.788E+01 | 0.107E+02 | 0.967E+01 |
| sp-dec | Pu | 240 | 94 | 146 | 98 | 38 | 172.000 | 194.880 | 0.193895E+17 | 0.134000E+17 | -0.60E+16 | -0.45E+00 | -0.160E+00 | 0.213E+04 | 0.263E+03 | -0.165E+02 | 0.328E+02 | 0.810E+01 | 0.107E+02 | 0.196E+02 |
| sp-dec | Pu | 242 | 94 | 148 | 99 | 38 | 178.600 | 200.900 | 0.887558E+11 | 0.677000E+11 | -0.21E+11 | -0.31E+00 | -0.118E+00 | 0.213E+04 | 0.258E+03 | -0.175E+02 | 0.284E+02 | 0.824E+01 | 0.107E+02 | 0.192E+02 |
| sp-dec | Pu | 244 | 94 | 150 | 102 | 38 | 177.200 | 200.200 | 0.892646E+11 | 0.660000E+11 | -0.23E+11 | -0.34E+00 | -0.131E+00 | 0.213E+04 | 0.166E+03 | 0.149E+02 | -0.391E+01 | 0.128E+02 | 0.107E+02 | 0.718E+02 |
| sp-dec | Cm | 242 | 96 | 146 | 96 | 38 | 181.100 | 210.800 | 0.756319E+07 | 0.700000E+07 | -0.56E+06 | -0.78E-01 | -0.336E-01 | 0.220E+04 | 0.219E+03 | -0.662E+01 | 0.135E+02 | 0.101E+02 | 0.107E+02 | 0.381E+02 |
| sp-dec | Cm | 244 | 96 | 148 | 98 | 38 | 181.800 | 210.100 | 0.196172E+08 | 0.132000E+08 | -0.64E+07 | -0.48E+00 | -0.172E+00 | 0.220E+04 | 0.215E+03 | -0.436E+01 | 0.117E+02 | 0.103E+02 | 0.107E+02 | 0.414E+02 |
| sp-dec | Cm | 244 | 96 | 148 | 98 | 38 | 185.500 | 200.760 | 0.182974E+08 | 0.134000E+08 | -0.49E+07 | -0.36E+00 | -0.135E+00 | 0.220E+04 | 0.268E+03 | -0.216E+02 | 0.288E+02 | 0.824E+01 | 0.107E+02 | 0.151E+02 |
| sp-dec | Cm | 248 | 96 | 152 | 99 | 38 | 182.000 | 207.800 | 0.195124E+07 | 0.415000E+07 | 0.22E+07 | 0.53E+00 | 0.328E+00 | 0.220E+04 | 0.227E+03 | -0.969E+01 | 0.160E+02 | 0.972E+01 | 0.107E+02 | 0.336E+02 |
| sp-dec | Cm | 250 | 96 | 154 | 98 | 38 | 182.300 | 198.180 | 0.238807E+05 | 0.200000E+05 | -0.39E+04 | -0.19E+00 | -0.770E-01 | 0.220E+04 | 0.268E+03 | -0.263E+02 | 0.307E+02 | 0.823E+01 | 0.107E+02 | 0.103E+02 |
| sp-dec | Cf | 246 | 98 | 148 | 102 | 40 | 195.600 | 209.025 | 0.268549E+04 | 0.180000E+04 | -0.89E+03 | -0.37E+00 | -0.174E+00 | 0.232E+04 | 0.276E+03 | -0.242E+02 | 0.277E+02 | 0.840E+01 | 0.107E+02 | 0.132E+02 |
| sp-dec | Cf | 246 | 98 | 148 | 94 | 39 | 195.600 | 209.030 | 0.244877E+04 | 0.200000E+04 | -0.45E+03 | -0.20E+00 | -0.879E-01 | 0.230E+04 | 0.259E+03 | -0.183E+02 | 0.216E+02 | 0.890E+01 | 0.107E+02 | 0.214E+02 |
| sp-dec | Cf | 250 | 98 | 152 | 100 | 39 | 185.100 | 220.500 | 0.115928E+05 | 0.113000E+05 | -0.29E+03 | -0.25E-01 | -0.111E-01 | 0.230E+04 | 0.172E+03 | 0.888E+01 | -0.481E+01 | 0.134E+02 | 0.107E+02 | 0.686E+02 |
| sp-dec | Cf | 252 | 98 | 154 | 101 | 39 | 184.100 | 219.100 | 0.106527E+03 | 0.860000E+02 | -0.21E+02 | -0.20E+00 | -0.930E-01 | 0.230E+04 | 0.167E+03 | 0.809E+01 | -0.606E+01 | 0.138E+02 | 0.107E+02 | 0.694E+02 |

```
sp-dec Cf   252  98 154 100 39 186.500 207.730 0.117539E+03 0.855000E+02 -0.32E+02 -0.31E+00 -0.138E+00 0.230E+04 0.260E+03 -0.242E+02 0.263E+02 0.884E+01 0.107E+02 0.151E+02
sp-dec Cf   254  98 156 102 39 186.100 206.720 0.409632E+00 0.166735E+00 -0.24E+00 -0.14E+01 -0.390E+00 0.230E+04 0.245E+03 -0.214E+02 0.210E+02 0.940E+01 0.107E+02 0.205E+02
sp-dec Cf   254  98 156 102 39 186.100 206.720 0.409632E+00 0.178000E+00 -0.23E+00 -0.13E+01 -0.362E+00 0.230E+04 0.245E+03 -0.214E+02 0.210E+02 0.940E+01 0.107E+02 0.205E+02
sp-dec Fm   256 100 156 102 40 196.900 234.700 0.380499E-03 0.330824E-03 -0.50E-04 -0.15E+00 -0.608E-01 0.240E+04 0.181E+03  0.200E+01 -0.542E+01 0.132E+02 0.107E+02 0.609E+02
sp-dec Fm   258 100 158 104 40 200.300 220.900 0.127911E-10 0.380000E-10  0.25E-10  0.65E+00  0.473E+00 0.240E+04 0.247E+03 -0.273E+02 0.164E+02 0.970E+01 0.107E+02 0.159E+02
al-dec TE   107  52  55   4  2   3.853   4.008 0.105947E-09 0.982331E-10 -0.77E-11 -0.59E-01 -0.328E-01 0.100E+03 0.151E+02 -0.254E+02 0.154E+02 0.663E+01 0.107E+02 0.413E+01
al-dec CS   114  55  59   4  2   3.239   3.357 0.485393E-07 0.180622E-07 -0.30E-07 -0.38E+00 -0.429E+00 0.106E+03 0.158E+02 -0.261E+02 0.188E+02 0.670E+01 0.107E+02 0.371E+01
al-dec TB   150  65  85   4  2   3.492   3.587 0.147994E-02 0.396988E-03 -0.11E-02 -0.49E+00 -0.571E+00 0.126E+03 0.179E+02 -0.240E+02 0.212E+02 0.702E+01 0.107E+02 0.727E+01
al-dec TB   151  65  86   4  2   3.409   3.497 0.850645E-03 0.200878E-02  0.12E-02  0.55E+00  0.373E+00 0.126E+03 0.178E+02 -0.250E+02 0.219E+02 0.706E+01 0.107E+02 0.649E+01
al-dec DY   151  66  85   4  2   4.069   4.180 0.662611E-05 0.340330E-04  0.27E-04  0.69E+00  0.711E+00 0.128E+03 0.182E+02 -0.242E+02 0.190E+02 0.704E+01 0.107E+02 0.715E+01
al-dec HO   152  67  85   4  2   4.454   4.507 0.158567E-05 0.158440E-05 -0.13E-08 -0.74E-03 -0.346E-03 0.130E+03 0.184E+02 -0.238E+02 0.180E+02 0.706E+01 0.107E+02 0.757E+01
al-dec HO   154  67  87   4  2   3.937   4.042 0.804562E-05 0.223591E-04  0.14E-04  0.25E+00  0.444E+00 0.130E+03 0.184E+02 -0.253E+02 0.202E+02 0.707E+01 0.107E+02 0.617E+01
al-dec ER   155  68  87   4  2   4.012   4.119 0.128590E-04 0.100768E-04 -0.28E-05 -0.18E+00 -0.106E+00 0.132E+03 0.186E+02 -0.250E+02 0.201E+02 0.708E+01 0.107E+02 0.656E+01
al-dec TM   154  69  85   4  2   5.031   5.094 0.119418E-06 0.104571E-06 -0.15E-07 -0.46E-01 -0.577E-01 0.134E+03 0.189E+02 -0.237E+02 0.168E+02 0.708E+01 0.107E+02 0.784E+01
al-dec TM   156  69  87   4  2   4.232   4.344 0.788374E-05 0.265546E-05 -0.52E-05 -0.16E+01 -0.473E+00 0.134E+03 0.189E+02 -0.246E+02 0.195E+02 0.709E+01 0.107E+02 0.691E+01
al-dec YB   155  70  85   4  2   5.200   5.337 0.473745E-07 0.568167E-07  0.94E-08  0.14E-01  0.789E-01 0.136E+03 0.192E+02 -0.239E+02 0.165E+02 0.709E+01 0.107E+02 0.770E+01
al-dec YB   157  70  87   4  2   4.504   4.622 0.248983E-05 0.122316E-05 -0.13E-05 -0.82E+00 -0.309E+00 0.136E+03 0.191E+02 -0.244E+02 0.188E+02 0.710E+01 0.107E+02 0.720E+01
al-dec LU   156  71  85   4  2   5.453   5.593 0.148925E-07 0.156539E-07  0.76E-09  0.47E-01  0.217E-01 0.138E+03 0.194E+02 -0.240E+02 0.162E+02 0.710E+01 0.107E+02 0.757E+01
al-dec LU   157  71  86   4  2   4.996   5.096 0.180585E-06 0.151786E-06 -0.29E-07 -0.54E-01 -0.754E-01 0.138E+03 0.193E+02 -0.242E+02 0.174E+02 0.714E+01 0.107E+02 0.759E+01
al-dec LU   158  71  87   4  2   4.669   4.790 0.216352E-05 0.335894E-06 -0.18E-05 -0.42E+01 -0.809E+00 0.138E+03 0.194E+02 -0.243E+02 0.186E+02 0.711E+01 0.107E+02 0.740E+01
al-dec HF   157  72  85   4  2   5.731   5.881 0.290477E-08 0.348569E-08  0.58E-09  0.16E+00  0.792E-01 0.140E+03 0.197E+02 -0.244E+02 0.158E+02 0.712E+01 0.107E+02 0.733E+01
al-dec HF   161  72  89   4  2   4.599   4.722 0.313534E-05 0.576723E-06 -0.26E-05 -0.31E+01 -0.735E+00 0.140E+03 0.196E+02 -0.248E+02 0.193E+02 0.715E+01 0.107E+02 0.704E+01
al-dec TA   159  73  86   4  2   5.600   5.662 0.590437E-08 0.164144E-07  0.11E-07  0.62E+00  0.444E+00 0.142E+03 0.198E+02 -0.246E+02 0.163E+02 0.719E+01 0.107E+02 0.742E+01
al-dec TA   160  73  87   4  2   5.412   5.446 0.490791E-07 0.491165E-07  0.37E-10  0.21E-03  0.331E-03 0.142E+03 0.198E+02 -0.242E+02 0.169E+02 0.718E+01 0.107E+02 0.770E+01
al-dec TA   161  73  88   4  2   5.148   5.280 0.158026E-06 0.915786E-07 -0.66E-07 -0.13E+00 -0.237E+00 0.142E+03 0.198E+02 -0.245E+02 0.177E+02 0.718E+01 0.107E+02 0.745E+01
al-dec W    159  74  85   4  2   6.295   6.443 0.987848E-10 0.259842E-09  0.16E-09  0.34E+00  0.420E+00 0.144E+03 0.201E+02 -0.254E+02 0.154E+02 0.717E+01 0.107E+02 0.650E+01
al-dec W    161  74  87   4  2   5.776   5.924 0.600869E-08 0.129921E-07  0.70E-08  0.49E+00  0.335E+00 0.144E+03 0.201E+02 -0.247E+02 0.165E+02 0.717E+01 0.107E+02 0.724E+01
al-dec RE   160  75  85   4  2   6.540   6.699 0.252479E-10 0.250336E-10 -0.21E-12 -0.40E-03 -0.370E-02 0.146E+03 0.203E+02 -0.261E+02 0.155E+02 0.718E+01 0.107E+02 0.594E+01
al-dec RE   165  75  90   4  2   5.512   5.658 0.524462E-07 0.652775E-07  0.13E-07  0.17E-01  0.950E-01 0.146E+03 0.202E+02 -0.247E+02 0.174E+02 0.723E+01 0.107E+02 0.748E+01
al-dec OS   165  76  89   4  2   6.183   6.317 0.164383E-08 0.224985E-08  0.61E-09  0.26E+00  0.136E+00 0.148E+03 0.205E+02 -0.249E+02 0.161E+02 0.722E+01 0.107E+02 0.720E+01
al-dec OS   167  76  91   4  2   5.839   5.979 0.217517E-07 0.265863E-07  0.48E-08  0.18E+00  0.872E-01 0.148E+03 0.205E+02 -0.246E+02 0.170E+02 0.722E+01 0.107E+02 0.752E+01
al-dec OS   169  76  93   4  2   5.577   5.717 0.974112E-07 0.107739E-06  0.10E-07  0.60E-01  0.438E-01 0.148E+03 0.205E+02 -0.248E+02 0.178E+02 0.723E+01 0.107E+02 0.742E+01
al-dec OS   171  76  95   4  2   5.244   5.370 0.429751E-06 0.263011E-06 -0.17E-06 -0.51E+00 -0.213E+00 0.148E+03 0.204E+02 -0.254E+02 0.190E+02 0.725E+01 0.107E+02 0.691E+01
al-dec OS   173  76  97   4  2   4.939   5.057 0.127887E-05 0.709813E-06 -0.57E-06 -0.57E+00 -0.256E+00 0.148E+03 0.204E+02 -0.263E+02 0.204E+02 0.727E+01 0.107E+02 0.609E+01
al-dec IR   165  77  88   4  2   6.715   6.800 0.207862E-10 0.950643E-11 -0.11E-10 -0.57E-01 -0.340E+00 0.150E+03 0.206E+02 -0.259E+02 0.152E+02 0.727E+01 0.107E+02 0.646E+01
al-dec IR   166  77  89   4  2   6.562   6.703 0.220175E-09 0.332725E-09  0.11E-09  0.11E+00  0.179E+00 0.150E+03 0.207E+02 -0.254E+02 0.157E+02 0.723E+01 0.107E+02 0.682E+01
al-dec IR   169  77  92   4  2   6.119   6.276 0.544387E-08 0.890435E-08  0.35E-08  0.33E-01  0.214E+00 0.150E+03 0.206E+02 -0.248E+02 0.165E+02 0.727E+01 0.107E+02 0.758E+01
al-dec IR   170  77  93   4  2   5.815   6.173 0.183625E-06 0.275686E-07 -0.16E-06 -0.38E+00 -0.824E+00 0.150E+03 0.205E+02 -0.239E+02 0.172E+02 0.733E+01 0.107E+02 0.871E+01
al-dec IR   172  77  95   4  2   5.828   5.991 0.516486E-07 0.693969E-07  0.18E-07  0.61E-01  0.128E+00 0.150E+03 0.207E+02 -0.248E+02 0.176E+02 0.726E+01 0.107E+02 0.747E+01
al-dec IR   173  77  96   4  2   5.672   5.840 0.657921E-07 0.684463E-07  0.27E-08  0.13E-01  0.172E-01 0.150E+03 0.206E+02 -0.252E+02 0.181E+02 0.730E+01 0.107E+02 0.725E+01
al-dec IR   174  77  97   4  2   5.478   5.624 0.232162E-06 0.158757E-06 -0.73E-07 -0.85E-01 -0.165E+00 0.150E+03 0.206E+02 -0.255E+02 0.188E+02 0.727E+01 0.107E+02 0.691E+01
al-dec IR   175  77  98   4  2   5.393   5.709 0.252202E-06 0.285193E-06  0.33E-07  0.95E-01  0.534E-01 0.150E+03 0.203E+02 -0.255E+02 0.189E+02 0.738E+01 0.107E+02 0.732E+01
al-dec IR   176  77  99   4  2   5.118   5.237 0.106389E-05 0.263011E-06 -0.80E-06 -0.18E+01 -0.607E+00 0.150E+03 0.206E+02 -0.264E+02 0.204E+02 0.729E+01 0.107E+02 0.604E+01
al-dec IR   177  77 100   4  2   5.011   5.127 0.104930E-05 0.950643E-06 -0.99E-07 -0.97E-01 -0.429E-01 0.150E+03 0.205E+02 -0.271E+02 0.211E+02 0.733E+01 0.107E+02 0.553E+01
al-dec PT   171  78  93   4  2   6.453   6.607 0.246920E-08 0.139428E-08 -0.11E-08 -0.67E+00 -0.248E+00 0.152E+03 0.209E+02 -0.247E+02 0.161E+02 0.727E+01 0.107E+02 0.761E+01
al-dec PT   173  78  95   4  2   6.211   6.353 0.139756E-07 0.121049E-07 -0.19E-08 -0.15E+00 -0.624E-01 0.152E+03 0.209E+02 -0.247E+02 0.168E+02 0.728E+01 0.107E+02 0.774E+01
al-dec PT   175  78  97   4  2   6.038   6.178 0.312397E-07 0.804877E-07  0.49E-07  0.18E+00  0.411E+00 0.152E+03 0.208E+02 -0.250E+02 0.175E+02 0.729E+01 0.107E+02 0.750E+01
al-dec PT   177  78  99   4  2   5.517   5.644 0.422317E-06 0.348569E-06 -0.74E-07 -0.19E+00 -0.834E-01 0.152E+03 0.208E+02 -0.256E+02 0.193E+02 0.731E+01 0.107E+02 0.690E+01
al-dec PT   179  78 101   4  2   5.195   5.395 0.120912E-05 0.668619E-06 -0.54E-06 -0.68E+00 -0.257E+00 0.152E+03 0.207E+02 -0.265E+02 0.206E+02 0.734E+01 0.107E+02 0.617E+01
```

```
al-dec PT    181  78 103   4 2   5.036   5.150  0.887610E-06 0.164778E-05  0.76E-06  0.32E+00  0.269E+00  0.152E+03  0.207E+02  -0.281E+02  0.221E+02  0.734E+01  0.107E+02  0.459E+01
al-dec PT    183  78 105   4 2   4.719   4.819  0.166344E-04 0.123584E-04 -0.43E-05 -0.14E+00 -0.129E+00  0.152E+03  0.206E+02  -0.291E+02  0.243E+02  0.736E+01  0.107E+02  0.373E+01
al-dec AU    170  79  91   4 2   7.107   7.170  0.198185E-10 0.196466E-10 -0.17E-12 -0.11E-02 -0.378E-02  0.154E+03  0.211E+02  -0.257E+02  0.150E+02  0.730E+01  0.107E+02  0.675E+01
al-dec AU    171  79  92   4 2   6.996   7.089  0.404875E-10 0.323218E-10 -0.82E-11 -0.23E-01 -0.978E-01  0.154E+03  0.210E+02  -0.255E+02  0.151E+02  0.732E+01  0.107E+02  0.708E+01
al-dec AU    172  79  93   4 2   6.870   6.923  0.244805E-09 0.243998E-09 -0.81E-12 -0.12E-02 -0.143E-02  0.154E+03  0.211E+02  -0.250E+02  0.154E+02  0.731E+01  0.107E+02  0.750E+01
al-dec AU    173  79  94   4 2   6.737   6.836  0.529389E-09 0.443633E-09 -0.86E-10 -0.12E+00 -0.768E-01  0.154E+03  0.210E+02  -0.250E+02  0.157E+02  0.732E+01  0.107E+02  0.763E+01
al-dec AU    174  79  95   4 2   6.629   6.699  0.245847E-08 0.516199E-08  0.27E-08  0.48E+00  0.322E+00  0.154E+03  0.211E+02  -0.247E+02  0.161E+02  0.730E+01  0.107E+02  0.786E+01
al-dec AU    175  79  96   4 2   6.439   6.562  0.582294E-08 0.507009E-08 -0.75E-09 -0.14E+00 -0.601E-01  0.154E+03  0.210E+02  -0.248E+02  0.165E+02  0.733E+01  0.107E+02  0.785E+01
al-dec AU    176  79  97   4 2   6.286   6.541  0.311407E-07 0.342231E-07  0.31E-08  0.54E-02  0.410E-01  0.154E+03  0.210E+02  -0.245E+02  0.170E+02  0.732E+01  0.107E+02  0.811E+01
al-dec AU    181  79 102   4 2   5.626   5.756  0.279703E-06 0.459477E-06  0.18E-06  0.32E+00  0.216E+00  0.154E+03  0.209E+02  -0.264E+02  0.198E+02  0.737E+01  0.107E+02  0.643E+01
al-dec AU    182  79 103   4 2   5.403   5.526  0.495732E-06 0.491165E-06 -0.46E-08 -0.74E-02 -0.402E-02  0.154E+03  0.210E+02  -0.271E+02  0.208E+02  0.735E+01  0.107E+02  0.557E+01
al-dec AU    183  79 104   4 2   5.346   5.466  0.533072E-06 0.135625E-05  0.82E-06  0.49E+00  0.406E+00  0.154E+03  0.208E+02  -0.277E+02  0.214E+02  0.739E+01  0.107E+02  0.518E+01
al-dec AU    184  79 105   4 2   5.187   5.232  0.362399E-05 0.150835E-05 -0.21E-05 -0.11E+01 -0.381E+00  0.154E+03  0.208E+02  -0.281E+02  0.227E+02  0.739E+01  0.107E+02  0.480E+01
al-dec AU    185  79 106   4 2   5.069   5.180  0.487791E-05 0.808046E-05  0.32E-05  0.17E+00  0.219E+00  0.154E+03  0.208E+02  -0.287E+02  0.234E+02  0.741E+01  0.107E+02  0.429E+01
al-dec AU    186  79 107   4 2   4.653   4.912  0.126532E-03 0.203438E-04 -0.11E-03 -0.36E+01 -0.794E+00  0.154E+03  0.207E+02  -0.289E+02  0.250E+02  0.745E+01  0.107E+02  0.427E+01
al-dec HG    173  80  93   4 2   7.203   7.380  0.381983E-10 0.221817E-10 -0.16E-10 -0.19E+00 -0.236E+00  0.156E+03  0.214E+02  -0.257E+02  0.152E+02  0.731E+01  0.107E+02  0.687E+01
al-dec HG    175  80  95   4 2   6.882   7.043  0.603939E-09 0.339063E-09 -0.26E-09 -0.56E+00 -0.251E+00  0.156E+03  0.213E+02  -0.250E+02  0.158E+02  0.731E+01  0.107E+02  0.752E+01
al-dec HG    177  80  97   4 2   6.579   6.740  0.723287E-08 0.403389E-08 -0.32E-08 -0.69E+00 -0.254E+00  0.156E+03  0.213E+02  -0.247E+02  0.166E+02  0.732E+01  0.107E+02  0.789E+01
al-dec HG    179  80  99   4 2   6.285   6.340  0.586515E-07 0.332725E-07 -0.25E-07 -0.20E+00 -0.246E+00  0.156E+03  0.212E+02  -0.248E+02  0.175E+02  0.736E+01  0.107E+02  0.799E+01
al-dec HG    183  80 103   4 2   5.904   6.039  0.817393E-07 0.297868E-06  0.22E-06  0.42E+00  0.562E+00  0.156E+03  0.212E+02  -0.264E+02  0.193E+02  0.736E+01  0.107E+02  0.642E+01
al-dec HG    185  80 105   4 2   5.653   5.774  0.352120E-06 0.155589E-05  0.12E-05  0.64E+00  0.645E+00  0.156E+03  0.211E+02  -0.272E+02  0.207E+02  0.738E+01  0.107E+02  0.572E+01
al-dec TL    177  81  96   4 2   6.907   7.067  0.375560E-09 0.570386E-09  0.19E-09  0.27E+00  0.181E+00  0.158E+03  0.215E+02  -0.254E+02  0.160E+02  0.736E+01  0.107E+02  0.735E+01
al-dec TL    179  81  98   4 2   6.567   6.718  0.734982E-08 0.728826E-08 -0.62E-10 -0.45E-03 -0.365E-02  0.158E+03  0.214E+02  -0.250E+02  0.169E+02  0.737E+01  0.107E+02  0.780E+01
al-dec TL    182  81 101   4 2   6.406   6.550  0.600546E-07 0.918955E-07  0.32E-07  0.13E+00  0.185E+00  0.158E+03  0.215E+02  -0.249E+02  0.177E+02  0.735E+01  0.107E+02  0.785E+01
al-dec PB    181  82  99   4 2   7.065   7.206  0.786996E-09 0.142596E-08  0.64E-09  0.31E+00  0.258E+00  0.160E+03  0.217E+02  -0.252E+02  0.161E+02  0.736E+01  0.107E+02  0.760E+01
al-dec PB    183  82 101   4 2   6.868   6.928  0.506731E-08 0.131506E-07  0.81E-08  0.59E+00  0.414E+00  0.160E+03  0.216E+02  -0.250E+02  0.167E+02  0.740E+01  0.107E+02  0.791E+01
al-dec PB    187  82 105   4 2   6.194   6.395  0.163382E-06 0.481659E-06  0.32E-06  0.55E+00  0.470E+00  0.160E+03  0.216E+02  -0.258E+02  0.190E+02  0.741E+01  0.107E+02  0.718E+01
al-dec BI    186  83 103   4 2   7.369   7.757  0.314456E-09 0.310543E-09 -0.39E-11 -0.89E-02 -0.544E-02  0.162E+03  0.218E+02  -0.253E+02  0.158E+02  0.745E+01  0.107E+02  0.786E+01
al-dec BI    187  83 104   4 2   7.721   7.778  0.432413E-10 0.918955E-11 -0.34E-10 -0.20E+00 -0.673E+00  0.162E+03  0.217E+02  -0.258E+02  0.155E+02  0.746E+01  0.107E+02  0.738E+01
al-dec BI    188  83 105   4 2   7.106   7.255  0.180001E-08 0.839734E-08  0.66E-08  0.76E+00  0.669E+00  0.162E+03  0.219E+02  -0.257E+02  0.170E+02  0.741E+01  0.107E+02  0.729E+01
al-dec BI    192  83 109   4 2   6.348   6.376  0.352001E-06 0.125485E-05  0.90E-06  0.65E+00  0.552E+00  0.162E+03  0.217E+02  -0.268E+02  0.203E+02  0.748E+01  0.107E+02  0.655E+01
al-dec BI    193  83 110   4 2   6.174   6.305  0.449373E-06 0.101402E-06 -0.35E-06 -0.12E+01 -0.647E+00  0.162E+03  0.216E+02  -0.273E+02  0.210E+02  0.748E+01  0.107E+02  0.600E+01
al-dec BI    194  83 111   4 2   5.892   5.918  0.915540E-05 0.364413E-05 -0.55E-05 -0.15E+01 -0.400E+00  0.162E+03  0.216E+02  -0.277E+02  0.226E+02  0.750E+01  0.107E+02  0.571E+01
al-dec BI    195  83 112   4 2   5.713   5.832  0.155506E-04 0.579892E-05 -0.98E-05 -0.16E+01 -0.428E+00  0.162E+03  0.216E+02  -0.283E+02  0.235E+02  0.750E+01  0.107E+02  0.512E+01
al-dec PO    189  84 105   4 2   7.532   7.701  0.145295E-09 0.110908E-09 -0.34E-10 -0.13E+00 -0.117E+00  0.164E+03  0.221E+02  -0.259E+02  0.161E+02  0.743E+01  0.107E+02  0.716E+01
al-dec PO    191  84 107   4 2   7.376   7.501  0.708519E-09 0.294699E-08  0.22E-08  0.74E+00  0.619E+00  0.164E+03  0.220E+02  -0.260E+02  0.168E+02  0.745E+01  0.107E+02  0.718E+01
al-dec PO    193  84 109   4 2   7.002   7.096  0.773830E-08 0.760514E-08 -0.13E-09 -0.34E-02 -0.754E-02  0.164E+03  0.220E+02  -0.264E+02  0.183E+02  0.747E+01  0.107E+02  0.686E+01
al-dec PO    195  84 111   4 2   6.699   6.746  0.628126E-07 0.608411E-07 -0.20E-08 -0.16E-01 -0.138E-01  0.164E+03  0.219E+02  -0.271E+02  0.199E+02  0.751E+01  0.107E+02  0.631E+01
al-dec PO    197  84 113   4 2   6.281   6.412  0.617634E-06 0.169848E-05  0.11E-05  0.54E+00  0.439E+00  0.164E+03  0.219E+02  -0.280E+02  0.218E+02  0.749E+01  0.107E+02  0.540E+01
al-dec PO    199  84 115   4 2   6.059   6.074  0.103644E-04 0.792836E-05 -0.24E-05 -0.14E+00 -0.116E+00  0.164E+03  0.217E+02  -0.290E+02  0.240E+02  0.756E+01  0.107E+02  0.470E+01
al-dec PO    213  84 129   4 2   8.376   8.537  0.137779E-12 0.133090E-12 -0.47E-14 -0.12E-01 -0.150E-01  0.164E+03  0.216E+02  -0.333E+02  0.205E+02  0.758E+01  0.107E+02  0.406E+00
al-dec AT    191  85 106   4 2   7.715   7.820  0.428446E-10 0.665450E-10  0.24E-10  0.15E+00  0.191E+00  0.166E+03  0.222E+02  -0.264E+02  0.160E+02  0.748E+01  0.107E+02  0.692E+01
al-dec AT    195  85 110   4 2   7.221   7.339  0.286612E-08 0.465815E-08  0.18E-08  0.37E+00  0.211E+00  0.166E+03  0.221E+02  -0.266E+02  0.181E+02  0.751E+01  0.107E+02  0.682E+01
al-dec AT    196  85 111   4 2   7.055   7.201  0.985347E-08 0.801709E-08 -0.18E-08 -0.22E+00 -0.896E-01  0.166E+03  0.222E+02  -0.267E+02  0.187E+02  0.748E+01  0.107E+02  0.659E+01
al-dec AT    197  85 112   4 2   6.959   7.100  0.147651E-07 0.122950E-07 -0.25E-08 -0.12E+00 -0.795E-01  0.166E+03  0.221E+02  -0.272E+02  0.194E+02  0.752E+01  0.107E+02  0.630E+01
al-dec AT    198  85 113   4 2   6.753   6.893  0.711072E-07 0.129921E-06  0.59E-07  0.26E+00  0.262E+00  0.166E+03  0.221E+02  -0.274E+02  0.203E+02  0.750E+01  0.107E+02  0.596E+01
al-dec AT    199  85 114   4 2   6.643   6.780  0.129285E-06 0.222767E-06  0.93E-07  0.13E+00  0.236E+00  0.166E+03  0.220E+02  -0.280E+02  0.211E+02  0.754E+01  0.107E+02  0.559E+01
al-dec AT    200  85 115   4 2   6.465   6.596  0.629555E-06 0.136259E-05  0.73E-06  0.53E+00  0.335E+00  0.166E+03  0.221E+02  -0.283E+02  0.221E+02  0.752E+01  0.107E+02  0.518E+01
al-dec AT    201  85 116   4 2   6.342   6.473  0.151144E-05 0.263011E-05  0.11E-05  0.42E+00  0.241E+00  0.166E+03  0.220E+02  -0.289E+02  0.230E+02  0.755E+01  0.107E+02  0.476E+01
```

```
al-dec AT  202 85 117  4 2  6.228  6.354 0.586877E-05 0.583061E-05 -0.38E-07 -0.65E-02 -0.283E-02 0.166E+03 0.220E+02 -0.292E+02 0.240E+02 0.753E+01 0.107E+02 0.434E+01
al-dec AT  203 85 118  4 2  6.087  6.210 0.196834E-04 0.140695E-04 -0.56E-05 -0.31E+00 -0.146E+00 0.166E+03 0.219E+02 -0.298E+02 0.251E+02 0.757E+01 0.107E+02 0.391E+01
al-dec AT  204 85 119  4 2  5.950  6.070 0.114255E-03 0.175298E-04 -0.97E-04 -0.23E+01 -0.814E+00 0.166E+03 0.220E+02 -0.301E+02 0.262E+02 0.755E+01 0.107E+02 0.349E+01
al-dec AT  205 85 120  4 2  5.902  6.019 0.175512E-03 0.511446E-04 -0.12E-03 -0.19E+01 -0.536E+00 0.166E+03 0.219E+02 -0.307E+02 0.269E+02 0.759E+01 0.107E+02 0.307E+01
al-dec AT  207 85 122  4 2  5.758  5.872 0.942607E-03 0.205339E-03 -0.74E-03 -0.11E+01 -0.662E+00 0.166E+03 0.218E+02 -0.315E+02 0.285E+02 0.760E+01 0.107E+02 0.230E+01
al-dec AT  209 85 124  4 2  5.647  5.757 0.262337E-02 0.617157E-03 -0.20E-02 -0.17E+01 -0.628E+00 0.166E+03 0.218E+02 -0.322E+02 0.297E+02 0.761E+01 0.107E+02 0.163E+01
al-dec AT  213 85 128  4 2  9.080  9.254 0.346887E-14 0.396101E-14  0.49E-15  0.12E+00  0.576E-01 0.166E+03 0.218E+02 -0.333E+02 0.188E+02 0.762E+01 0.107E+02 0.669E+00
al-dec AT  214 85 129  4 2  8.819  8.987 0.233298E-13 0.176820E-13 -0.56E-14 -0.31E+00 -0.120E+00 0.166E+03 0.219E+02 -0.333E+02 0.197E+02 0.759E+01 0.107E+02 0.514E+00
al-dec AT  215 85 130  4 2  8.026  8.178 0.165112E-11 0.316881E-11  0.15E-11  0.23E-01  0.283E-01 0.166E+03 0.218E+02 -0.336E+02 0.218E+02 0.763E+01 0.107E+02 0.389E+00
al-dec AT  216 85 131  4 2  7.802  7.947 0.131900E-10 0.950643E-11 -0.37E-11 -0.35E-01 -0.142E+00 0.166E+03 0.219E+02 -0.335E+02 0.226E+02 0.760E+01 0.107E+02 0.290E+00
al-dec AT  217 85 132  4 2  7.067  7.201 0.166040E-08 0.102353E-08 -0.64E-09 -0.55E+00 -0.210E+00 0.166E+03 0.218E+02 -0.337E+02 0.250E+02 0.763E+01 0.107E+02 0.213E+00
al-dec AT  218 85 133  4 2  6.756  6.874 0.414226E-07 0.475321E-07  0.61E-08  0.43E-01  0.598E-01 0.166E+03 0.219E+02 -0.337E+02 0.263E+02 0.759E+01 0.107E+02 0.155E+00
al-dec RN  195 86 109  4 2  7.555  7.690 0.291972E-09 0.158440E-09 -0.13E-09 -0.60E+00 -0.265E+00 0.168E+03 0.224E+02 -0.265E+02 0.170E+02 0.748E+01 0.107E+02 0.679E+01
al-dec RN  197 86 111  4 2  7.357  7.410 0.246997E-08 0.665450E-09 -0.18E-08 -0.22E+01 -0.570E+00 0.168E+03 0.223E+02 -0.267E+02 0.181E+02 0.752E+01 0.107E+02 0.681E+01
al-dec RN  199 86 113  4 2  6.989  7.130 0.218347E-07 0.186960E-07 -0.31E-08 -0.28E-01 -0.674E-01 0.168E+03 0.224E+02 -0.271E+02 0.195E+02 0.751E+01 0.107E+02 0.631E+01
al-dec RN  201 86 115  4 2  6.752  6.861 0.159553E-06 0.224985E-06  0.65E-07  0.14E+00  0.149E+00 0.168E+03 0.223E+02 -0.278E+02 0.210E+02 0.754E+01 0.107E+02 0.572E+01
al-dec RN  203 86 117  4 2  6.499  6.630 0.113997E-05 0.139428E-05  0.25E-06  0.17E+00  0.875E-01 0.168E+03 0.223E+02 -0.287E+02 0.227E+02 0.755E+01 0.107E+02 0.493E+01
al-dec RN  205 86 119  4 2  6.268  6.390 0.115959E-04 0.538697E-05 -0.62E-05 -0.11E+01 -0.333E+00 0.168E+03 0.222E+02 -0.296E+02 0.247E+02 0.756E+01 0.107E+02 0.408E+01
al-dec RN  207 86 121  4 2  6.131  6.251 0.627025E-04 0.175869E-04 -0.45E-04 -0.91E+00 -0.552E+00 0.168E+03 0.222E+02 -0.305E+02 0.263E+02 0.758E+01 0.107E+02 0.321E+01
al-dec RN  209 86 123  4 2  6.039  6.155 0.203379E-03 0.547570E-04 -0.15E-03 -0.21E+01 -0.570E+00 0.168E+03 0.221E+02 -0.314E+02 0.277E+02 0.759E+01 0.107E+02 0.238E+01
al-dec RN  211 86 125  4 2  5.852  5.964 0.120798E-02 0.166553E-02  0.46E-03  0.24E+00  0.139E+00 0.168E+03 0.221E+02 -0.322E+02 0.293E+02 0.760E+01 0.107E+02 0.165E+01
al-dec RN  215 86 129  4 2  8.674  8.839 0.127141E-12 0.728826E-13 -0.54E-13 -0.14E+00 -0.242E+00 0.168E+03 0.221E+02 -0.333E+02 0.204E+02 0.762E+01 0.107E+02 0.650E+00
al-dec RN  219 86 133  4 2  6.819  6.946 0.500355E-07 0.125485E-06  0.75E-07  0.48E+00  0.399E+00 0.168E+03 0.220E+02 -0.337E+02 0.264E+02 0.762E+01 0.107E+02 0.198E+00
al-dec RN  221 86 135  4 2  6.037  6.148 0.741317E-04 0.488630E-04 -0.25E-04 -0.43E+00 -0.181E+00 0.168E+03 0.221E+02 -0.338E+02 0.297E+02 0.762E+01 0.107E+02 0.101E+00
al-dec FR  200 87 113  4 2  7.471  7.621 0.169646E-08 0.155272E-08 -0.14E-09 -0.85E-01 -0.385E-01 0.170E+03 0.226E+02 -0.270E+02 0.183E+02 0.752E+01 0.107E+02 0.643E+01
al-dec FR  201 87 114  4 2  7.361  7.516 0.316206E-08 0.218648E-08 -0.98E-09 -0.38E+00 -0.160E+00 0.170E+03 0.225E+02 -0.274E+02 0.189E+02 0.756E+01 0.107E+02 0.627E+01
al-dec FR  202 87 115  4 2  7.241  7.389 0.107157E-07 0.950643E-08 -0.12E-08 -0.71E-02 -0.520E-01 0.170E+03 0.226E+02 -0.275E+02 0.195E+02 0.753E+01 0.107E+02 0.604E+01
al-dec FR  203 87 116  4 2  7.131  7.260 0.226577E-07 0.173968E-07 -0.53E-08 -0.11E-01 -0.115E+00 0.170E+03 0.224E+02 -0.280E+02 0.203E+02 0.758E+01 0.107E+02 0.576E+01
al-dec FR  204 87 117  4 2  7.031  7.171 0.595488E-07 0.602074E-07  0.66E-09  0.30E-02  0.478E-02 0.170E+03 0.225E+02 -0.282E+02 0.210E+02 0.755E+01 0.107E+02 0.542E+01
al-dec FR  205 87 118  4 2  6.915  7.055 0.123034E-06 0.124217E-06  0.12E-08  0.47E-02  0.416E-02 0.170E+03 0.224E+02 -0.288E+02 0.218E+02 0.759E+01 0.107E+02 0.504E+01
al-dec FR  206 87 119  4 2  6.792  6.923 0.450375E-06 0.507009E-06  0.57E-07  0.11E+00  0.514E-01 0.170E+03 0.225E+02 -0.291E+02 0.227E+02 0.757E+01 0.107E+02 0.462E+01
al-dec FR  207 87 120  4 2  6.767  6.900 0.496731E-06 0.468984E-06 -0.28E-07 -0.55E-01 -0.250E-01 0.170E+03 0.223E+02 -0.297E+02 0.234E+02 0.761E+01 0.107E+02 0.419E+01
al-dec FR  208 87 121  4 2  6.641  6.772 0.194221E-05 0.187277E-05 -0.69E-07 -0.35E-01 -0.158E-01 0.170E+03 0.224E+02 -0.300E+02 0.243E+02 0.758E+01 0.107E+02 0.373E+01
al-dec FR  209 87 122  4 2  6.646  6.777 0.156529E-05 0.160025E-05  0.35E-07  0.19E-01  0.959E-02 0.170E+03 0.223E+02 -0.307E+02 0.248E+02 0.762E+01 0.107E+02 0.328E+01
al-dec FR  210 87 123  4 2  6.545  6.672 0.543621E-05 0.604609E-05  0.61E-06  0.36E-01  0.462E-01 0.170E+03 0.224E+02 -0.310E+02 0.257E+02 0.760E+01 0.107E+02 0.283E+01
al-dec FR  211 87 124  4 2  6.534  6.660 0.403546E-05 0.589398E-05  0.19E-05  0.19E+00  0.165E+00 0.170E+03 0.223E+02 -0.316E+02 0.262E+02 0.763E+01 0.107E+02 0.241E+01
al-dec FR  212 87 125  4 2  6.406  6.529 0.181499E-04 0.380257E-04  0.20E-04  0.41E+00  0.321E+00 0.170E+03 0.223E+02 -0.319E+02 0.271E+02 0.761E+01 0.107E+02 0.201E+01
al-dec FR  213 87 126  4 2  6.775  6.905 0.401670E-06 0.109641E-05  0.69E-06  0.58E+00  0.436E+00 0.170E+03 0.222E+02 -0.324E+02 0.260E+02 0.765E+01 0.107E+02 0.164E+01
al-dec FR  215 87 128  4 2  9.360  9.540 0.476437E-14 0.272518E-14 -0.20E-14 -0.71E+00 -0.243E+00 0.170E+03 0.222E+02 -0.330E+02 0.187E+02 0.765E+01 0.107E+02 0.105E+01
al-dec FR  216 87 129  4 2  9.004  9.174 0.482878E-13 0.221817E-13 -0.26E-13 -0.31E+00 -0.338E+00 0.170E+03 0.223E+02 -0.331E+02 0.198E+02 0.763E+01 0.107E+02 0.814E+00
al-dec FR  217 87 130  4 2  8.315  8.469 0.146466E-11 0.602074E-12 -0.86E-12 -0.12E+01 -0.386E+00 0.170E+03 0.222E+02 -0.335E+02 0.217E+02 0.766E+01 0.107E+02 0.622E+00
al-dec FR  218 87 131  4 2  7.867  8.014 0.481902E-10 0.316881E-10 -0.17E-10 -0.74E-01 -0.182E+00 0.170E+03 0.223E+02 -0.335E+02 0.232E+02 0.763E+01 0.107E+02 0.468E+00
al-dec FR  219 87 132  4 2  7.312  7.448 0.131835E-08 0.633762E-09 -0.68E-09 -0.98E+00 -0.318E+00 0.170E+03 0.222E+02 -0.338E+02 0.249E+02 0.767E+01 0.107E+02 0.347E+00
al-dec FR  220 87 133  4 2  6.677  6.801 0.453424E-06 0.868254E-06  0.41E-06  0.43E+00  0.282E+00 0.170E+03 0.223E+02 -0.337E+02 0.274E+02 0.764E+01 0.107E+02 0.253E+00
al-dec FR  221 87 134  4 2  6.341  6.458 0.472481E-05 0.931630E-05  0.46E-05  0.35E+00  0.295E+00 0.170E+03 0.222E+02 -0.340E+02 0.286E+02 0.767E+01 0.107E+02 0.183E+00
al-dec RA  203 88 115  4 2  7.612  7.730 0.127764E-08 0.760514E-09 -0.52E-09 -0.58E+00 -0.225E+00 0.172E+03 0.228E+02 -0.274E+02 0.185E+02 0.756E+01 0.107E+02 0.620E+01
al-dec RA  205 88 117  4 2  7.370  7.490 0.901327E-08 0.538697E-08 -0.36E-08 -0.54E+00 -0.224E+00 0.172E+03 0.227E+02 -0.279E+02 0.199E+02 0.757E+01 0.107E+02 0.577E+01
al-dec RA  209 88 121  4 2  7.003  7.140 0.160889E-06 0.148934E-06 -0.12E-07 -0.56E-01 -0.335E-01 0.172E+03 0.226E+02 -0.296E+02 0.228E+02 0.760E+01 0.107E+02 0.422E+01
al-dec RA  211 88 123  4 2  6.910  7.046 0.382148E-06 0.411945E-06  0.30E-07  0.63E-01  0.326E-01 0.172E+03 0.226E+02 -0.306E+02 0.242E+02 0.761E+01 0.107E+02 0.328E+01
```

```
al-dec RA    213  88 125   4 2   6.730   6.860  0.172953E-05  0.520952E-05   0.35E-05   0.21E+00   0.479E+00   0.172E+03   0.226E+02  -0.316E+02   0.258E+02   0.763E+01   0.107E+02   0.238E+01
al-dec RA    217  88 129   4 2   8.992   9.161  0.109819E-12  0.507009E-13  -0.59E-13  -0.52E+00  -0.336E+00   0.172E+03   0.225E+02  -0.330E+02   0.201E+02   0.765E+01   0.107E+02   0.100E+01
al-dec RA    219  88 131   4 2   7.988   8.138  0.446716E-10  0.316881E-09   0.27E-09   0.66E+00   0.851E+00   0.172E+03   0.225E+02  -0.335E+02   0.231E+02   0.765E+01   0.107E+02   0.586E+00
al-dec RA    221  88 133   4 2   6.754   6.884  0.520842E-06  0.887266E-06   0.37E-06   0.39E+00   0.231E+00   0.172E+03   0.225E+02  -0.338E+02   0.275E+02   0.766E+01   0.107E+02   0.322E+00
al-dec RA    225  88 137   4 2   5.006   5.789  0.904740E-01  0.407940E-01  -0.50E-01  -0.11E+01  -0.346E+00   0.172E+03   0.216E+02  -0.351E+02   0.341E+02   0.796E+01   0.107E+02   0.296E+00
al-dec AC    206  89 117   4 2   7.790   7.945  0.886557E-09  0.697138E-09  -0.19E-09  -0.19E+00  -0.104E+00   0.174E+03   0.230E+02  -0.278E+02   0.187E+02   0.757E+01   0.107E+02   0.593E+01
al-dec AC    208  89 119   4 2   7.572   7.730  0.477907E-08  0.301037E-08  -0.18E-08  -0.47E+00  -0.201E+00   0.174E+03   0.229E+02  -0.284E+02   0.200E+02   0.759E+01   0.107E+02   0.541E+01
al-dec AC    209  89 120   4 2   7.577   7.725  0.440099E-08  0.310543E-08  -0.13E-08  -0.33E+00  -0.151E+00   0.174E+03   0.228E+02  -0.289E+02   0.205E+02   0.763E+01   0.107E+02   0.506E+01
al-dec AC    210  89 121   4 2   7.462   7.607  0.123209E-07  0.110908E-07  -0.12E-08  -0.87E-02  -0.457E-01   0.174E+03   0.229E+02  -0.292E+02   0.213E+02   0.760E+01   0.107E+02   0.464E+01
al-dec AC    211  89 122   4 2   7.480   7.620  0.904820E-08  0.792202E-08  -0.11E-08  -0.67E-02  -0.577E-01   0.174E+03   0.228E+02  -0.298E+02   0.218E+02   0.764E+01   0.107E+02   0.419E+01
al-dec AC    212  89 123   4 2   7.379   7.519  0.227999E-07  0.294699E-07   0.67E-08   0.35E-01   0.111E+00   0.174E+03   0.228E+02  -0.302E+02   0.226E+02   0.762E+01   0.107E+02   0.371E+01
al-dec AC    213  89 124   4 2   7.364   7.503  0.183723E-07  0.253505E-07   0.70E-08   0.38E-01   0.140E+00   0.174E+03   0.227E+02  -0.309E+02   0.231E+02   0.766E+01   0.107E+02   0.323E+01
al-dec AC    214  89 125   4 2   7.214   7.350  0.727531E-07  0.259842E-06   0.19E-06   0.58E+00   0.553E+00   0.174E+03   0.228E+02  -0.312E+02   0.241E+02   0.763E+01   0.107E+02   0.275E+01
al-dec AC    215  89 126   4 2   7.604   7.744  0.272336E-08  0.538697E-08   0.27E-08   0.71E-01   0.296E+00   0.174E+03   0.227E+02  -0.318E+02   0.233E+02   0.767E+01   0.107E+02   0.231E+01
al-dec AC    217  89 128   4 2   9.650   9.832  0.499233E-14  0.218648E-14  -0.28E-14  -0.12E+01  -0.359E+00   0.174E+03   0.226E+02  -0.327E+02   0.184E+02   0.768E+01   0.107E+02   0.153E+01
al-dec AC    218  89 129   4 2   9.205   9.380  0.761838E-13  0.342231E-13  -0.42E-13  -0.13E+00  -0.348E+00   0.174E+03   0.227E+02  -0.329E+02   0.198E+02   0.766E+01   0.107E+02   0.121E+01
al-dec AC    219  89 130   4 2   8.664   8.830  0.887762E-12  0.373919E-12  -0.51E-12  -0.60E+00  -0.376E+00   0.174E+03   0.226E+02  -0.333E+02   0.213E+02   0.769E+01   0.107E+02   0.942E+00
al-dec AC    221  89 132   4 2   7.645   7.783  0.623681E-09  0.164778E-08   0.10E-08   0.60E+00   0.422E+00   0.174E+03   0.226E+02  -0.337E+02   0.245E+02   0.770E+01   0.107E+02   0.545E+00
al-dec AC    222  89 133   4 2   7.009   7.134  0.161424E-06  0.158440E-06  -0.30E-08  -0.94E-02  -0.810E-02   0.174E+03   0.227E+02  -0.338E+02   0.270E+02   0.767E+01   0.107E+02   0.405E+00
al-dec AC    223  89 134   4 2   6.662   6.783  0.152479E-05  0.399270E-05   0.25E-05   0.18E+00   0.418E+00   0.174E+03   0.226E+02  -0.340E+02   0.282E+02   0.771E+01   0.107E+02   0.297E+00
al-dec AC    224  89 135   4 2   6.214   6.357  0.212652E-03  0.317134E-03   0.10E-03   0.47E-01   0.174E+00   0.174E+03   0.227E+02  -0.339E+02   0.303E+02   0.767E+01   0.107E+02   0.214E+00
al-dec TH    209  90 119   4 2   8.080   8.238  0.245761E-09  0.120415E-09  -0.13E-09  -0.21E+00  -0.310E+00   0.176E+03   0.231E+02  -0.282E+02   0.186E+02   0.760E+01   0.107E+02   0.564E+01
al-dec TH    213  90 123   4 2   7.692   7.839  0.289392E-08  0.443633E-08   0.15E-08   0.30E+00   0.186E+00   0.176E+03   0.231E+02  -0.299E+02   0.214E+02   0.763E+01   0.107E+02   0.409E+01
al-dec TH    215  90 125   4 2   7.520   7.665  0.829126E-08  0.380257E-07   0.30E-07   0.29E+00   0.661E+00   0.176E+03   0.230E+02  -0.309E+02   0.229E+02   0.765E+01   0.107E+02   0.311E+01
al-dec TH    219  90 129   4 2   9.340   9.510  0.547106E-13  0.332725E-13  -0.21E-13  -0.17E+00  -0.216E+00   0.176E+03   0.229E+02  -0.327E+02   0.195E+02   0.767E+01   0.107E+02   0.143E+01
al-dec TH    221  90 131   4 2   8.470   8.628  0.791084E-11  0.548204E-10   0.47E-10   0.31E+00   0.841E+00   0.176E+03   0.229E+02  -0.333E+02   0.222E+02   0.768E+01   0.107E+02   0.871E+00
al-dec TH    223  90 133   4 2   7.435   7.567  0.103491E-07  0.190129E-07   0.87E-08   0.11E+00   0.264E+00   0.176E+03   0.229E+02  -0.338E+02   0.258E+02   0.769E+01   0.107E+02   0.499E+00
al-dec TH    225  90 135   4 2   6.797   6.920  0.186012E-05  0.165792E-04   0.15E-04   0.61E+00   0.950E+00   0.176E+03   0.229E+02  -0.340E+02   0.283E+02   0.770E+01   0.107E+02   0.270E+00
al-dec PA    212  91 121   4 2   8.270   8.429  0.144038E-09  0.161609E-09   0.18E-10   0.23E-01   0.500E-01   0.178E+03   0.234E+02  -0.287E+02   0.189E+02   0.762E+01   0.107E+02   0.523E+01
al-dec PA    216  91 125   4 2   7.948   8.097  0.705052E-09  0.529191E-08   0.46E-08   0.76E+00   0.875E+00   0.178E+03   0.233E+02  -0.306E+02   0.215E+02   0.765E+01   0.107E+02   0.344E+01
al-dec PA    217  91 126   4 2   8.336   8.489  0.411986E-10  0.114077E-09   0.73E-10   0.20E+00   0.442E+00   0.178E+03   0.231E+02  -0.313E+02   0.209E+02   0.769E+01   0.107E+02   0.295E+01
al-dec PA    224  91 133   4 2   7.528   7.694  0.121287E-07  0.250336E-07   0.13E-07   0.60E-01   0.315E+00   0.178E+03   0.231E+02  -0.337E+02   0.258E+02   0.770E+01   0.107E+02   0.600E+00
al-dec PA    225  91 134   4 2   7.245   7.380  0.538697E-07  0.538697E-07  -0.38E-08  -0.44E-01  -0.297E-01   0.178E+03   0.230E+02  -0.340E+02   0.268E+02   0.774E+01   0.107E+02   0.451E+00
al-dec PA    226  91 135   4 2   6.864   6.987  0.270643E-05  0.342231E-05   0.72E-06   0.99E-01   0.102E+00   0.178E+03   0.231E+02  -0.340E+02   0.284E+02   0.771E+01   0.107E+02   0.334E+00
al-dec PA    227  91 136   4 2   6.466   6.580  0.509422E-04  0.728192E-04   0.22E-04   0.28E+00   0.155E+00   0.178E+03   0.230E+02  -0.342E+02   0.299E+02   0.774E+01   0.107E+02   0.244E+00
al-dec PA    228  91 137   4 2   6.126   6.264  0.328342E-02  0.250970E-02  -0.77E-03  -0.30E+00  -0.117E+00   0.178E+03   0.231E+02  -0.341E+02   0.317E+02   0.771E+01   0.107E+02   0.176E+00
al-dec U     217  92 125   4 2   8.018   8.160  0.407127E-09  0.507009E-09   0.10E-09   0.14E+00   0.953E-01   0.180E+03   0.235E+02  -0.304E+02   0.210E+02   0.767E+01   0.107E+02   0.373E+01
al-dec U     223  92 131   4 2   8.780   8.940  0.254577E-11  0.570386E-12  -0.20E-11  -0.27E+01  -0.650E+00   0.180E+03   0.233E+02  -0.331E+02   0.215E+02   0.771E+01   0.107E+02   0.119E+01
al-dec U     227  92 135   4 2   7.060   7.211  0.928665E-06  0.209141E-05   0.12E-05   0.29E+00   0.353E+00   0.180E+03   0.233E+02  -0.340E+02   0.280E+02   0.773E+01   0.107E+02   0.402E+00
al-dec U     229  92 137   4 2   6.360   6.475  0.607079E-03  0.110275E-03  -0.50E-03  -0.43E+01  -0.741E+00   0.180E+03   0.233E+02  -0.342E+02   0.310E+02   0.774E+01   0.107E+02   0.218E+00
al-dec U     233  92 141   4 2   4.824   4.909  0.712004E+05  0.159200E+06   0.88E+05   0.55E+00   0.349E+00   0.180E+03   0.233E+02  -0.344E+02   0.392E+02   0.774E+01   0.107E+02   0.901E-01
al-dec NP    226  93 133   4 2   8.060   8.200  0.531293E-09  0.110908E-08   0.58E-09   0.40E+00   0.320E+00   0.182E+03   0.236E+02  -0.336E+02   0.243E+02   0.773E+01   0.107E+02   0.818E+00
al-dec NP    229  93 136   4 2   6.893   7.010  0.462053E-05  0.760514E-05   0.30E-05   0.20E+00   0.216E+00   0.182E+03   0.234E+02  -0.343E+02   0.289E+02   0.777E+01   0.107E+02   0.356E+00
al-dec PU    239  94 145   4 2   5.157   5.244  0.100455E+05  0.241100E+05   0.14E+05   0.58E+00   0.380E+00   0.184E+03   0.237E+02  -0.346E+02   0.386E+02   0.778E+01   0.107E+02   0.730E-01
al-dec AM    242  95 147   4 2   5.518   5.588  0.316700E+03  0.141000E+03  -0.18E+03  -0.12E+01  -0.351E+00   0.186E+03   0.238E+02  -0.347E+02   0.372E+02   0.780E+01   0.107E+02   0.591E-01
al-dec CM    241  96 145   4 2   6.081   6.185  0.849696E+00  0.898015E-01  -0.76E+00  -0.80E+01  -0.976E+00   0.188E+03   0.240E+02  -0.347E+02   0.346E+02   0.782E+01   0.107E+02   0.967E-01
al-dec BK    243  97 146   4 2   6.761   6.874  0.113398E-02  0.513347E-03  -0.62E-03  -0.83E+00  -0.344E+00   0.190E+03   0.242E+02  -0.349E+02   0.319E+02   0.786E+01   0.107E+02   0.115E+00
al-dec BK    245  97 148   4 2   6.348   6.455  0.926250E-01  0.135250E-01  -0.79E-01  -0.36E+01  -0.836E+00   0.190E+03   0.242E+02  -0.349E+02   0.339E+02   0.786E+01   0.107E+02   0.851E-01
al-dec CF    243  98 145   4 2   7.170   7.330  0.107310E-03  0.203438E-04  -0.87E-04  -0.29E+01  -0.722E+00   0.192E+03   0.245E+02  -0.347E+02   0.308E+02   0.784E+01   0.107E+02   0.199E+00
```

```
al-dec  CF      245  98 147   4  2   7.138    7.258  0.163053E-03  0.882196E-04  -0.75E-04  -0.80E+00  -0.267E+00  0.192E+03  0.244E+02  -0.348E+02  0.310E+02  0.785E+01  0.107E+02  0.143E+00
al-dec  ES      245  99 146   4  2   7.780    7.909  0.797252E-06  0.209141E-05   0.13E-05   0.32E+00   0.419E+00  0.194E+03  0.246E+02  -0.349E+02  0.288E+02  0.789E+01  0.107E+02  0.270E+00
al-dec  ES      247  99 148   4  2   7.323    7.490  0.511092E-04  0.865085E-05  -0.42E-04  -0.74E+00  -0.771E+00  0.194E+03  0.246E+02  -0.350E+02  0.307E+02  0.789E+01  0.107E+02  0.185E+00
al-dec  ES      253  99 154   4  2   6.633    6.739  0.618801E-01  0.560438E-01  -0.58E-02  -0.91E-01  -0.430E-01  0.194E+03  0.246E+02  -0.351E+02  0.339E+02  0.790E+01  0.107E+02  0.575E-01
al-dec  ES      255  99 156   4  2   6.401    6.436  0.488248E+00  0.108966E+00  -0.38E+00  -0.27E+01  -0.651E+00  0.194E+03  0.245E+02  -0.353E+02  0.349E+02  0.793E+01  0.107E+02  0.397E-01
al-dec  FM      251 100 151   4  2   7.306    7.425  0.299288E-03  0.604609E-03   0.31E-03   0.20E+00   0.305E+00  0.196E+03  0.248E+02  -0.350E+02  0.315E+02  0.789E+01  0.107E+02  0.166E+00
al-dec  FM      253 100 153   4  2   7.083    7.197  0.276510E-02  0.821355E-02   0.54E-02   0.13E+00   0.473E+00  0.196E+03  0.248E+02  -0.350E+02  0.325E+02  0.789E+01  0.107E+02  0.108E+00
al-dec  FM      255 100 155   4  2   7.127    7.240  0.230391E-02  0.228953E-02  -0.14E-04  -0.47E-02  -0.272E-02  0.196E+03  0.248E+02  -0.351E+02  0.324E+02  0.789E+01  0.107E+02  0.693E-01
al-dec  FM      257 100 157   4  2   6.752    6.864  0.131683E+00  0.275154E+00   0.14E+00   0.51E+00   0.320E+00  0.196E+03  0.248E+02  -0.351E+02  0.342E+02  0.789E+01  0.107E+02  0.443E-01
al-dec  MD      250 101 149   4  2   7.831    8.310  0.173887E-05  0.164778E-05  -0.91E-07  -0.50E-01  -0.234E-01  0.198E+03  0.248E+02  -0.350E+02  0.293E+02  0.798E+01  0.107E+02  0.494E+00
al-dec  MD      255 101 154   4  2   7.752    7.906  0.106842E-04  0.513347E-04   0.41E-04   0.74E+00   0.682E+00  0.198E+03  0.250E+02  -0.352E+02  0.302E+02  0.794E+01  0.107E+02  0.141E+00
al-dec  MD      256 101 155   4  2   7.733    7.856  0.255196E-04  0.146399E-03   0.12E-03   0.80E+00   0.759E+00  0.198E+03  0.250E+02  -0.351E+02  0.305E+02  0.791E+01  0.107E+02  0.112E+00
al-dec  MD      257 101 156   4  2   7.440    7.558  0.175087E-03  0.629706E-03   0.45E-03   0.38E+00   0.556E+00  0.198E+03  0.249E+02  -0.353E+02  0.315E+02  0.794E+01  0.107E+02  0.878E-01
al-dec  NO      252 102 150   4  2   8.415    8.549  0.169668E-06  0.719320E-07  -0.98E-07  -0.19E+00  -0.373E+00  0.200E+03  0.252E+02  -0.348E+02  0.280E+02  0.794E+01  0.107E+02  0.613E+00
al-dec  NO      253 102 151   4  2   8.144    8.400  0.213938E-05  0.308008E-05   0.94E-06   0.29E-01   0.158E+00  0.200E+03  0.252E+02  -0.348E+02  0.291E+02  0.792E+01  0.107E+02  0.496E+00
al-dec  NO      256 102 154   4  2   8.448    8.581  0.991535E-07  0.922123E-07  -0.69E-08  -0.28E-01  -0.315E-01  0.200E+03  0.252E+02  -0.352E+02  0.282E+02  0.795E+01  0.107E+02  0.245E+00
al-dec  NO      259 102 157   4  2   7.685    7.890  0.111782E-03  0.110275E-03  -0.15E-05  -0.13E-01  -0.590E-02  0.200E+03  0.252E+02  -0.352E+02  0.312E+02  0.792E+01  0.107E+02  0.116E+00
al-dec  LR      257 103 154   4  2   8.861    9.010  0.124562E-07  0.204705E-07   0.80E-08   0.99E-02   0.216E+00  0.202E+03  0.253E+02  -0.351E+02  0.272E+02  0.797E+01  0.107E+02  0.433E+00
al-dec  LR      258 103 155   4  2   8.680    8.900  0.101785E-06  0.129921E-06   0.28E-07   0.13E+00   0.106E+00  0.202E+03  0.255E+02  -0.350E+02  0.280E+02  0.794E+01  0.107E+02  0.339E+00
al-dec  RF      257 104 153   4  2   8.903    9.044  0.849157E-07  0.123584E-06   0.39E-07   0.15E+00   0.163E+00  0.204E+03  0.256E+02  -0.345E+02  0.274E+02  0.796E+01  0.107E+02  0.944E+00
al-dec  RF      259 104 155   4  2   8.770    9.120  0.108444E-06  0.101402E-06  -0.70E-08  -0.24E-01  -0.292E-01  0.204E+03  0.256E+02  -0.349E+02  0.279E+02  0.798E+01  0.107E+02  0.613E+00
al-dec  DB      257 105 152   4  2   9.074    9.230  0.179141E-06  0.475321E-07  -0.13E-06  -0.20E+00  -0.576E+00  0.206E+03  0.258E+02  -0.337E+02  0.269E+02  0.800E+01  0.107E+02  0.191E+01
al-dec  DB      260 105 155   4  2   9.121    9.371  0.656099E-07  0.481659E-07  -0.17E-07  -0.38E-01  -0.134E+00  0.206E+03  0.259E+02  -0.345E+02  0.273E+02  0.797E+01  0.107E+02  0.100E+01
al-dec  DB      262 105 157   4  2   8.670    9.005  0.770718E-06  0.110908E-05   0.34E-06   0.27E+00   0.158E+00  0.206E+03  0.258E+02  -0.349E+02  0.288E+02  0.799E+01  0.107E+02  0.639E+00
al-dec  DS      271 110 161   4  2  10.738   10.899  0.149730E-09  0.516516E-10  -0.98E-10  -0.68E-01  -0.462E+00  0.216E+03  0.268E+02  -0.344E+02  0.246E+02  0.807E+01  0.107E+02  0.152E+01
cl-dec  Fr      221  87 134  14  6  29.280   31.283  0.158942E+08  0.930000E+07  -0.66E+07  -0.64E+00  -0.233E+00  0.486E+03  0.815E+02  -0.265E+02  0.337E+02  0.596E+01  0.107E+02  0.846E-01
cl-dec  Ra      221  88 133  14  6  30.340   32.393  0.561406E+06  0.470000E+06  -0.91E+05  -0.16E+00  -0.772E-01  0.492E+03  0.826E+02  -0.263E+02  0.321E+02  0.595E+01  0.107E+02  0.188E+00
cl-dec  Ac      225  89 136  14  6  28.570   30.467  0.162317E+10  0.460000E+10   0.30E+10   0.53E+00   0.452E+00  0.498E+03  0.831E+02  -0.266E+02  0.358E+02  0.599E+01  0.107E+02  0.954E-01
cl-dec  Pa      231  91 140  23  9  46.680   51.840  0.295840E+19  0.300000E+19   0.42E+17   0.10E-01   0.606E-02  0.738E+03  0.120E+03  -0.272E+02  0.457E+02  0.617E+01  0.107E+02  0.206E+00
cl-dec  Pa      231  91 140  24 10  54.140   60.419  0.380943E+16  0.250000E+16  -0.13E+16  -0.37E+00  -0.183E+00  0.810E+03  0.128E+03  -0.280E+02  0.436E+02  0.632E+01  0.107E+02  0.140E+00
cl-dec  U       233  92 141  25 10  54.320   60.837  0.224465E+18  0.220000E+18  -0.45E+16  -0.14E-01  -0.873E-02  0.820E+03  0.130E+03  -0.278E+02  0.451E+02  0.628E+01  0.107E+02  0.218E+00
cl-dec  U       235  92 143  24 10  51.500   57.358  0.971691E+20  0.900000E+20  -0.72E+19  -0.72E-01  -0.333E-01  0.820E+03  0.130E+03  -0.280E+02  0.479E+02  0.629E+01  0.107E+02  0.640E-01
cl-dec  Np      237  93 144  30 12  65.520   75.017  0.232911E+20  0.160000E+21   0.14E+21   0.53E+00   0.837E+00  0.972E+03  0.149E+03  -0.289E+02  0.483E+02  0.655E+01  0.107E+02  0.194E+00
cl-dec  Am      241  95 146  34 14  80.600   93.838  0.664526E+18  0.670000E+18   0.55E+16   0.71E-02   0.356E-02  0.113E+04  0.170E+03  -0.296E+02  0.474E+02  0.667E+01  0.107E+02  0.692E-01
sp-dec  Pu      239  94 145 100 41 177.900  184.300  0.105703E+17  0.800000E+16  -0.26E+16  -0.26E+00  -0.121E+00  0.217E+04  0.304E+03  -0.289E+02  0.449E+02  0.714E+01  0.107E+02  0.289E+01
sp-dec  Fm      257 100 157 103 40 197.100  236.300  0.153407E+03  0.131100E+03  -0.22E+02  -0.97E-01  -0.682E-01  0.240E+04  0.176E+03   0.957E+01  -0.738E+01  0.136E+02  0.107E+02  0.703E+02
pr-rad  I       109  53  56   1  1   0.813    0.829  0.452044E-11  0.326387E-11  -0.13E-11  -0.38E+00  -0.141E+00  0.520E+02  0.695E+01  -0.331E+02  0.218E+02  0.748E+01  0.107E+02  0.160E+00
pr-rad  Cs      113  55  58   1  1   0.960    0.978  0.234687E-12  0.529191E-12   0.29E-12   0.53E+00   0.353E+00  0.540E+02  0.719E+01  -0.333E+02  0.207E+02  0.752E+01  0.107E+02  0.139E+00
pr-rad  La      117  57  60   1  1   0.806    0.823  0.774321E-09  0.744670E-09  -0.30E-10  -0.39E-01  -0.170E-01  0.560E+02  0.742E+01  -0.335E+02  0.244E+02  0.755E+01  0.107E+02  0.116E+00
pr-rad  Ho      141  67  74   1  1   1.169    1.190  0.139436E-09  0.129921E-09  -0.95E-11  -0.68E-01  -0.307E-01  0.660E+02  0.851E+01  -0.344E+02  0.246E+02  0.776E+01  0.107E+02  0.919E-01
pr-rad  Lu      150  71  79   1  1   1.261    1.283  0.848064E-09  0.133090E-08   0.48E-09   0.36E+00   0.196E+00  0.700E+02  0.899E+01  -0.344E+02  0.253E+02  0.779E+01  0.107E+02  0.280E+00
pr-rad  Ir      167  77  90   1  1   1.064    1.086  0.213410E-08  0.111542E-08  -0.10E-08  -0.91E+00  -0.282E+00  0.760E+02  0.954E+01  -0.338E+02  0.251E+02  0.796E+01  0.107E+02  0.169E+01
pr-rad  Au      171  79  92   1  1   1.444    1.469  0.271843E-12  0.538697E-12   0.27E-12   0.47E+00   0.297E+00  0.780E+02  0.975E+01  -0.336E+02  0.211E+02  0.800E+01  0.107E+02  0.198E+01
pr-rad  Tl      177  81  96   1  1   1.156    1.180  0.575315E-09  0.570386E-09  -0.49E-11  -0.84E-02  -0.374E-02  0.800E+02  0.995E+01  -0.331E+02  0.238E+02  0.804E+01  0.107E+02  0.272E+01
pr-rad  Bi      185  83 102   1  1   1.598    1.624  0.159675E-11  0.158440E-11  -0.12E-13  -0.77E-02  -0.337E-02  0.820E+02  0.101E+02  -0.324E+02  0.206E+02  0.810E+01  0.107E+02  0.365E+01
al-deY 118294   294 118 176   4  2  11.660   11.820  0.105715E-10  0.218648E-10   0.11E-10   0.27E+00   0.316E+00  0.232E+03  0.282E+02  -0.364E+02  0.254E+02  0.824E+01  0.107E+02  0.314E+00
al-deY 117294   294 117 177   4  2  10.810   11.180  0.180245E-08  0.161609E-08  -0.19E-09  -0.66E-01  -0.474E-01  0.230E+03  0.281E+02  -0.363E+02  0.276E+02  0.819E+01  0.107E+02  0.130E+00
al-deY 117293   293 117 176   4  2  10.600   11.320  0.290767E-09  0.697138E-09   0.41E-09   0.43E+00   0.380E+00  0.230E+03  0.276E+02  -0.369E+02  0.274E+02  0.833E+01  0.107E+02  0.197E+00
al-deY 116293   293 116 177   4  2  10.560   10.710  0.361975E-08  0.180622E-08  -0.18E-08  -0.57E+00  -0.302E+00  0.228E+03  0.279E+02  -0.364E+02  0.279E+02  0.818E+01  0.107E+02  0.636E-01
```

```
al-deY 116292 292 116 176  4 2 10.630  10.780 0.128962E-08 0.411945E-09 -0.88E-09 -0.14E+01 -0.496E+00  0.228E+03  0.278E+02 -0.364E+02  0.276E+02  0.820E+01  0.107E+02  0.820E-01
al-deY 119291 291 116 175  4 2 10.740  10.890 0.949886E-09 0.602074E-09 -0.35E-09 -0.30E+00 -0.198E+00  0.228E+03  0.279E+02 -0.363E+02  0.273E+02  0.818E+01  0.107E+02  0.105E+00
al-deY 116290 290 116 174  4 2 10.850  11.000 0.274688E-09 0.263011E-09 -0.12E-10 -0.31E-01 -0.189E-01  0.228E+03  0.278E+02 -0.364E+02  0.268E+02  0.820E+01  0.107E+02  0.134E+00
al-deY 115290 290 115 175  4 2  9.780  10.410 0.576848E-07 0.205973E-07 -0.37E-07 -0.10E+01 -0.447E+00  0.226E+03  0.274E+02 -0.366E+02  0.294E+02  0.825E+01  0.107E+02  0.119E+00
al-deY 115289 289 115 174  4 2 10.150  10.490 0.155258E-07 0.104571E-07 -0.51E-08 -0.36E+00 -0.172E+00  0.226E+03  0.276E+02 -0.363E+02  0.285E+02  0.819E+01  0.107E+02  0.132E+00
al-deY 115288 288 115 173  4 2 10.290  10.490 0.137537E-07 0.519685E-08 -0.86E-08 -0.14E+01 -0.423E+00  0.226E+03  0.277E+02 -0.361E+02  0.283E+02  0.815E+01  0.107E+02  0.166E+00
al-deY 115287 287 115 172  4 2 10.610  10.760 0.611894E-09 0.117246E-08  0.56E-09  0.22E+00  0.282E+00  0.226E+03  0.276E+02 -0.363E+02  0.270E+02  0.819E+01  0.107E+02  0.211E+00
al-deY 114289 289 114 175  4 2  9.840   9.980 0.133746E-06 0.602074E-07 -0.74E-07 -0.89E+00 -0.347E+00  0.224E+03  0.275E+02 -0.362E+02  0.293E+02  0.815E+01  0.107E+02  0.926E-01
al-deY 114288 288 114 174  4 2  9.930  10.070 0.398991E-07 0.209141E-07 -0.19E-07 -0.75E+00 -0.281E+00  0.224E+03  0.274E+02 -0.363E+02  0.289E+02  0.817E+01  0.107E+02  0.121E+00
al-deY 114287 287 114 173  4 2 10.030  10.170 0.306147E-07 0.152103E-07 -0.15E-07 -0.79E+00 -0.304E+00  0.224E+03  0.275E+02 -0.361E+02  0.286E+02  0.815E+01  0.107E+02  0.156E+00
al-deY 114286 286 114 172  4 2 10.210  10.350 0.527629E-08 0.380257E-08 -0.15E-08 -0.29E+00 -0.142E+00  0.224E+03  0.274E+02 -0.362E+02  0.279E+02  0.817E+01  0.107E+02  0.201E+00
al-deY 113286 286 113 173  4 2  9.610   9.790 0.442662E-06 0.301037E-06 -0.14E-06 -0.28E+00 -0.167E+00  0.222E+03  0.274E+02 -0.360E+02  0.296E+02  0.812E+01  0.107E+02  0.138E+00
al-deY 113285 285 113 172  4 2  9.470  10.010 0.118227E-06 0.133090E-06  0.15E-07  0.84E-01  0.514E-01  0.222E+03  0.270E+02 -0.364E+02  0.295E+02  0.822E+01  0.107E+02  0.202E+00
al-deY 113284 284 113 171  4 2  9.100  10.120 0.424175E-07 0.288362E-07 -0.14E-07 -0.40E+00 -0.168E+00  0.222E+03  0.264E+02 -0.370E+02  0.296E+02  0.841E+01  0.107E+02  0.431E+00
al-deY 113283 283 113 170  4 2 10.230  10.380 0.223980E-08 0.237661E-08  0.14E-09  0.20E-01  0.257E-01  0.222E+03  0.272E+02 -0.360E+02  0.274E+02  0.816E+01  0.107E+02  0.300E+00
al-deY 113282 282 113 169  4 2 10.630  10.780 0.360111E-09 0.231323E-08  0.20E-08  0.30E+00  0.808E+00  0.222E+03  0.273E+02 -0.358E+02  0.263E+02  0.812E+01  0.107E+02  0.382E+00
al-deY 112285 285 112 173  4 2  9.190   9.320 0.386660E-05 0.887266E-06 -0.30E-05 -0.25E+01 -0.639E+00  0.220E+03  0.271E+02 -0.360E+02  0.306E+02  0.811E+01  0.107E+02  0.116E+00
al-deY 112283 283 112 171  4 2  9.530   9.660 0.263708E-06 0.133090E-06 -0.13E-06 -0.78E+00 -0.297E+00  0.220E+03  0.271E+02 -0.359E+02  0.293E+02  0.811E+01  0.107E+02  0.200E+00
al-deY 112281 281 112 169  4 2 10.310  10.460 0.124886E-08 0.316881E-08  0.19E-08  0.11E+00  0.404E+00  0.220E+03  0.271E+02 -0.358E+02  0.269E+02  0.811E+01  0.107E+02  0.336E+00
al-deY 111282 282 111 171  4 2  8.860  10.530 0.289512E-05 0.316881E-05  0.27E-06  0.51E-01  0.392E-01  0.218E+03  0.264E+02 -0.359E+02  0.303E+02  0.826E+01  0.107E+02  0.896E+00
al-deY 111281 281 111 170  4 2  9.280   9.410 0.432422E-06 0.538697E-06  0.11E-06  0.15E+00  0.954E-01  0.218E+03  0.268E+02 -0.360E+02  0.296E+02  0.812E+01  0.107E+02  0.211E+00
al-deY 111280 280 111 169  4 2  9.510  10.030 0.552981E-07 0.145765E-06  0.90E-07  0.53E+00  0.421E+00  0.218E+03  0.268E+02 -0.359E+02  0.287E+02  0.814E+01  0.107E+02  0.305E+00
al-deY 111278 278 111 167  4 2 10.690  10.850 0.723194E-10 0.133090E-09  0.61E-10  0.16E+00  0.265E+00  0.218E+03  0.270E+02 -0.355E+02  0.254E+02  0.809E+01  0.107E+02  0.462E+00
al-deY 110279 279 110 169  4 2  9.710   9.850 0.163728E-07 0.665450E-08 -0.97E-08 -0.12E+01 -0.391E+00  0.216E+03  0.267E+02 -0.358E+02  0.280E+02  0.808E+01  0.107E+02  0.211E+00
al-deY 110277 277 110 167  4 2 10.570  10.720 0.560123E-10 0.190129E-09  0.13E-09  0.13E+00  0.531E+00  0.216E+03  0.267E+02 -0.358E+02  0.254E+02  0.808E+01  0.107E+02  0.364E+00
al-deY 109278 278 109 169  4 2  9.380   9.580 0.116824E-06 0.142596E-06  0.26E-07  0.10E+00  0.866E-01  0.214E+03  0.266E+02 -0.357E+02  0.287E+02  0.804E+01  0.107E+02  0.149E+00
al-deY 109276 276 109 167  4 2  9.170  10.030 0.460625E-08 0.142596E-07  0.97E-08  0.53E+00  0.491E+00  0.214E+03  0.259E+02 -0.364E+02  0.281E+02  0.827E+01  0.107E+02  0.399E+00
al-deY 109275 275 109 166  4 2 10.330  10.480 0.682534E-10 0.633762E-09  0.57E-09  0.41E+00  0.968E+00  0.214E+03  0.265E+02 -0.357E+02  0.255E+02  0.809E+01  0.107E+02  0.347E+00
al-deY 108273 273 108 165  4 2  9.590   9.730 0.823271E-08 0.633762E-08 -0.19E-08 -0.43E-01 -0.114E+00  0.212E+03  0.264E+02 -0.355E+02  0.274E+02  0.804E+01  0.107E+02  0.313E+00
al-deY 107274 274 107 167  4 2  8.730   8.840 0.232947E-05 0.139428E-05 -0.94E-06 -0.37E+00 -0.223E+00  0.210E+03  0.262E+02 -0.356E+02  0.300E+02  0.802E+01  0.107E+02  0.111E+00
al-deY 107272 272 107 165  4 2  8.550   9.180 0.548800E-06 0.345400E-06 -0.20E-06 -0.50E+00 -0.201E+00  0.210E+03  0.258E+02 -0.360E+02  0.298E+02  0.815E+01  0.107E+02  0.251E+00
al-deY 107271 271 107 164  4 2  9.280   9.420 0.174127E-07 0.475321E-07  0.30E-07  0.22E+00  0.436E+00  0.210E+03  0.261E+02 -0.356E+02  0.278E+02  0.805E+01  0.107E+02  0.265E+00
al-deY 107270 270 107 163  4 2  8.930   9.060 0.522486E-06 0.193297E-05  0.14E-05  0.13E+00  0.568E+00  0.210E+03  0.262E+02 -0.354E+02  0.291E+02  0.802E+01  0.107E+02  0.349E+00
al-deY 106271 271 106 165  4 2  8.540   8.670 0.404656E-05 0.304206E-05 -0.10E-05 -0.17E+00 -0.124E+00  0.208E+03  0.260E+02 -0.355E+02  0.301E+02  0.800E+01  0.107E+02  0.119E+00
al-deY 106269 269 106 163  4 2  8.570   8.700 0.287035E-05 0.380257E-05  0.93E-06  0.41E-01  0.122E+00  0.208E+03  0.260E+02 -0.354E+02  0.299E+02  0.801E+01  0.107E+02  0.213E+00
al-deY 112282 282 112 170  4 2 10.805  10.960 0.289211E-10 0.288362E-10 -0.85E-13 -0.22E-02 -0.128E-02  0.220E+03  0.270E+02 -0.360E+02  0.254E+02  0.813E+01  0.107E+02  0.260E+00
```